\documentclass[fleqn,usenatbib]{mnras}
\usepackage{comment}
\usepackage{graphicx}	
\usepackage{amsmath}	
\usepackage{amssymb}	
\usepackage{longtable}
\usepackage{ragged2e}
\usepackage{verbatim}

\title[\emph{NuSTAR} and \emph{Swift} observations of 4U~1901+03]{Evolution of timing and spectral characteristics of 4U~1901+03 during its 2019 outburst using the \emph{Swift} and \emph{NuSTAR} observatories}

\author[Aru Beri  et.al. ]{Aru Beri$^{1,2}$ \thanks{a.beri@soton.ac.uk}, 
 Tinku$^{1}$, 
 Nirmal K. Iyer$^{3,4}$, 
 Chandreyee Maitra$^{5}$ \\
 $^1$Indian Institute of Science Education and Research (IISER) Mohali, 
Punjab 140306, India  \\
$^{2}$Physics \& Astronomy, University of Southampton, Southampton, Hampshire SO17 1BJ, UK \\
$^{3}$ KTH Royal Institute of Technology, Department of Physics, 106 91 Stockholm, Sweden \\
$^{4}$ The Oskar Klein Centre for Cosmoparticle Physics, AlbaNova University Centre, 106 91 Stockholm, Sweden \\
 $^{5}$ Max-Planck-Institut für extraterrestrische Physik, Giessenbachstraße 1, 85748 Garching, Germany \\
}

\begin{document}
%
%
%
\maketitle

\label{firstpage}
\begin{abstract}

We report the results from a detailed timing and spectral study of a transient X-ray
	pulsar, 4U~1901+03 during its 2019 outburst.~We performed broadband spectroscopy in the 1-70~$\rm keV$ energy band using four observations made with \emph{Swift} and \emph{NuSTAR} 
	at different intensity levels.~Our timing results reveal the presence of highly variable pulse profiles dependent on both luminosity and energy.~Our spectroscopy results showed the presence of a cyclotron resonance scattering feature~(CRSF) at $\sim$~30~${\rm keV}$.~This feature at 30~${\rm keV}$ is highly luminosity and pulse-phase dependent.~Phase-averaged spectra during the last two observations, made close to the declining phase of the outburst showed the presence of this feature at around $30~\rm{keV}$.~The existence of CRSF at 30~${\rm keV}$ during these observations is well supported by an abrupt change in the shape of pulse profiles found close to this energy.~We also found that 30~${\rm keV}$ feature was 
	significantly detected in the pulse-phase resolved spectra of observations made at relatively high luminosities.~Moreover,~all spectral fit parameters 
	showed a strong pulse phase dependence.~In line with the previous findings,~an absorption feature at around $10~\rm{keV}$ is significantly observed in the phase-averaged X-ray spectra of all observations and also showed a strong pulse phase dependence.

\end{abstract}


%
%
%
%

\section{Introduction}\label{sec:intro}

A majority of accretion powered X-ray pulsars belong to the class of High Mass X-ray Binaries where the companion star
is a Be type star or an OB type supergiant.~Be-X-ray binaries~(BeXRBs from now) are mostly transient systems that go into type I outburst close to the periastron passage, and giant type II outbursts which are rare \citep[][and references therein]{Reig11}. Transient X-ray pulsars are excellent candidates to understand changes in the accretion geometry and corresponding changes in timing and spectral signatures due to magnetically-driven accretion \citep[see e.g.,][]{Parmar89, Devasia11,Maitra12}. \\

4U~1901+03~(1901 from now) is a transient X-ray pulsar that was discovered with the \emph{Uhuru} and \emph{Vela 5B} observatories during its outburst in 1970-1971 \citep{Forman76,Priedhorsky84}. This source was classified as a BeXRB based only on its X-ray timing properties. The pulse period~($P_{spin}$) of 2.76 seconds \citep{Galloway05} and the orbital period~($P_{orb}$) of 22.6 days \citep{Galloway05,Jenke11,Tuo2020} placed it well in the region populated by BeXRBs of the $P_{orb}-P_{spin}$ diagram \citep{Corbet86}.~This source was not detected in X-rays until February 2003, when the source underwent a second giant outburst that lasted for about five months \citep{Galloway05}.~During its 2003 outburst, the peak X-ray luminosity
observed was $\sim$~$10^{38}$ $\rm{ergs~s^{-1}}$ that changed by almost three orders of magnitude to $10^{35}$ $\rm{ergs~s^{-1}}$, assuming a distance of 10~\rm{kpc} \citep{Galloway05}.
Later,~this source was detected in outburst in 2011, however, this outburst was much shorter~(1 month duration) and was fainter compared to its 2003 outburst with the peak X-ray luminosity of about {\bf{$7 {\times}10^{36}$ $\rm{ergs~s^{-1}}$,}}~assuming source distance of 10~\rm{kpc} \citep{Sootome11,Reig16}. 
On February 8, 2019, 1901 was again detected in outburst with \emph{MAXI}--\textsc{GSC}. This was recorded as the fourth outburst from this source in 40 years \citep{Nakajima19,Mereminskiy19,Hemphill19}, and lasted for about $\sim$~160~$days$.~During this outburst several optical and X-ray observations were made and the peak X-ray luminosity measured during the 2019 outburst is about {\bf{$1.4~{\times}~10^{37}$ $\rm{ergs~s^{-1}}$,}} assuming source distance of 3~\rm{kpc} \citep{Ji2020}. Based on optical spectroscopy of the candidate Be star counterpart~(B8/9 IV star), the source distance was proposed to be greater than 12~kpc \citep{Strader19, McCollum19}.~\citet{Tuo2020} using  \emph{Insight}-\emph{Hard X-ray Modulation Telescope}~(\emph{Insight-HXMT}) data derived a similar value of
the source distance.
However,~this is in contrast to that measured with \emph{Gaia} which suggests 1901 to be a nearby source at around $3.0^{+2.0}_{-1.1}~\rm{kpc}$ \citep{Balier-Jones18}. \\

The timing and spectral properties of 1901 were studied extensively during its previous bright outburst in 2003. \citep{Galloway05,Chen08,Lei09,James11, Reig16}. The pulse profiles showed complex morphology that were luminosity dependent \citep{Lei09, James11}.~At high luminosity (above $10^{37}\rm{ergs~s^{-1}}$), it showed a double-peak structure
that changed to a single peak towards the end of the outburst.~A quasi-periodic oscillation~(QPO) at 0.135~Hz was also detected in its light curves during 2003 outburst \citep{James11}.
A similar pulse profile evolution has also been found with the \emph{NICER} and \emph{Insight-HXMT} observations made during the 2019 outburst of 1901. These observations also revealed critical luminosity to be $\sim$~$10^{37}~\rm{ergs~s^{-1}}$ \citep{Ji2020,Tuo2020}.
\citet{Tuo2020} used value of the critical luminosity to estimate magnetic field of the neutron star in this system. Based on
the accretion torque model, the authors suggested the value to be $\sim$~$4\times{10^{12}}~\rm{Gauss}$. \\

The continuum X-ray spectra, observed with \emph{Rossi X-ray Timing Explorer}~(\emph{RXTE}) was well modelled using an
 absorbed power law and exponential cut-off or a model consisting of thermal Comptonization component. A significant excess above 10 keV was also seen, indicating the presence
of cyclotron resonance scattering feature~(CRSF) in its X-ray spectra \citep{Molkov03, Galloway05, James11, Reig16}.
~A pulse-phase resolved spectroscopy carried out by \citet{Lei09} during the 2003 outburst indicated that the main pulse peak has the hardest spectrum. This is a common characteristic of an accreting X-ray pulsar \citep[see e.g.][]{Kreykenbohm99,Maitra_2013}.~Moreover, 10~keV feature observed in the X-ray spectra of 1901 was also found to be 
strongly dependent on the X-ray flux and on the pulse phase \citep{Reig16}. \\

During the 2019 outburst of 1901, \emph{Swift} and \emph{NuSTAR} observations were made that allowed to study its broadband X-ray spectra (1--70~keV).~The X-ray continuum was described with a combination of an absorbed blackbody and a power law with exponential cut-off \citep{Mereminskiy19,Coley19}. The iron emission line between 6 and 7~keV was also found in its X-ray spectrum \citep{Mereminskiy19,Jaisawal19,Coley19}, in addition to the 10~keV feature \citep{Mereminskiy19, Coley19}.~Moreover, one of the \emph{NuSTAR} observations made close to the declining phase of the outburst
revealed the presence of negative residuals in the X-ray spectra around 30~\rm{keV} and this narrow absorption feature was interpreted as a CRSF \citep{Coley19}. \\

Here in this paper, we report a detailed timing and spectral study of 1901 during its 2019 outburst.~This paper is organised as follows.~The observational data and procedure are described in the following section~(Section~2).~In Section~3, we describe the timing analysis while Section~4 is focused on the details on spectral analysis.~In Section~5, we discuss the timing and spectral results obtained.

\begin{figure}
\includegraphics[height=0.55\columnwidth]{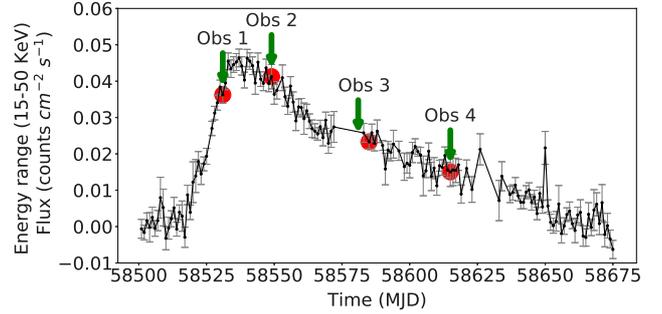}
\caption{This plot shows the \emph{Swift}-\textsc{BAT} lightcurve of 4U~1901+03 during its 2019 outburst. The red dots and green arrows indicate dates during which the \emph{NuSTAR} and \emph{Swift} observations were made respectively.}
\label{outburst}
\end{figure}

\begin{table*}
\centering
	
	\caption{Observations made with {\emph NuSTAR} and {\emph Swift} during the 2019 outburst of 4U~1901+03.} 
	\label{tab:obs}
	\scriptsize{\begin{tabular}{|c|c|l|c|c|}
	\hline        
	
	          &     &   {\emph{NuSTAR}} &  &     \\
	\hline
	Obs~ID	  &	Exp Time (ks) &	Start Time (MJD)    &  Pulse Period (s) \\
	\hline \hline

		90501305001~(Obs~1)  &	 $\sim$~44   &	58531.121	&   $2.76288\pm0.00003$             \\ 
		90502307002~(Obs~2) &	 $\sim$~38   & 	58549.308	&  $2.76154\pm0.00003$               \\ 
		90502307004~(Obs~3) &	 $\sim$~54   &  58584.946	&  $2.76211\pm0.00004$            \\ 
		90501324002~(Obs~4)  &	 $\sim$~102  &	58615.752	&  $2.76075\pm0.00003$          \\ 

   \hline \hline
	
   &     &  {\emph{Swift}} &   &    \\
   
   \hline
	Obs ID &	Exp Time (ks) &	Start Time (MJD)  & Pulse Period~(s)  \\
	\hline \hline
		00088846001~(Obs~1)  &	 $\sim$~7    & 	58531.768  &   $2.7630\pm0.0001$ \\ 
		00088849001~(Obs~2) &	 $\sim$~12   &  58549.426  &   $2.7615 \pm0.0001$   \\ 
		00037044015~(Obs~3)  &	 $\sim$~6.5  &  58581.052   &  $2.7628 \pm0.0001$    \\ 
		00088870001~(Obs~4) &	 $\sim$~6.5  &	58615.774  &  $2.7607\pm0.0002$   \\ 
   \hline \hline
  	\end{tabular}}  
	 \flushleft
Note :  The Gaussian 1-$\sigma$
uncertainties reported for the pulse periods were determined using the widths of the fitted Gaussian distributions to `efsearch' results. \\
\end{table*}

\begin{figure*}
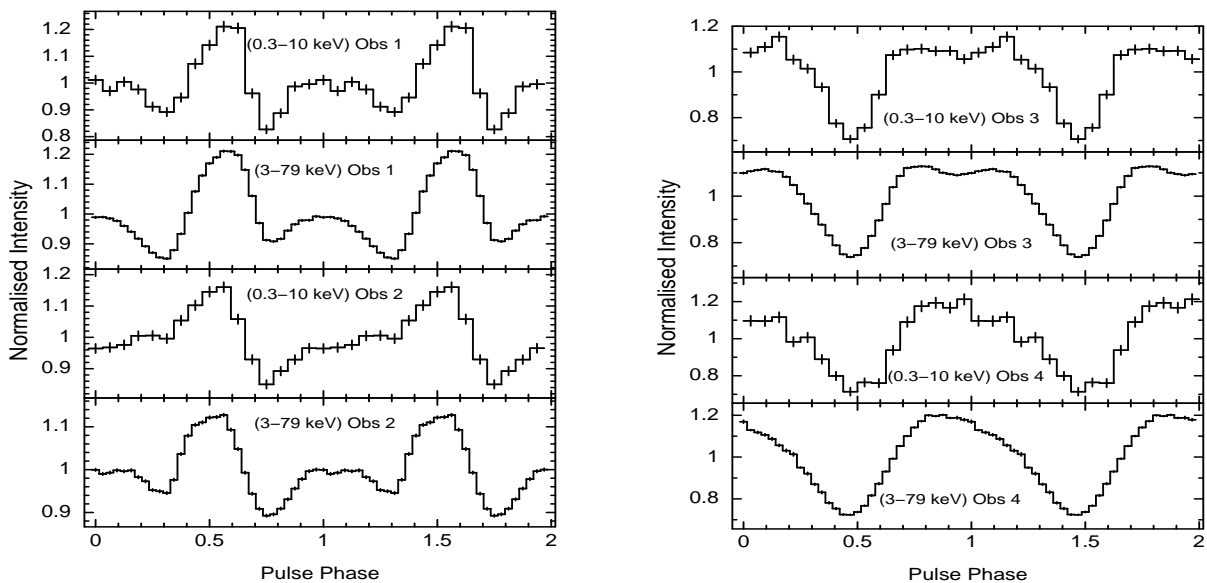

\centering
\begin{minipage}{0.45\textwidth}
\includegraphics[height=\columnwidth,width=\columnwidth,angle=0]{fig2.eps}
\end{minipage}
\hspace{0.02\linewidth}
\begin{minipage}{0.45\textwidth}
\includegraphics[height=\columnwidth,width=\columnwidth,angle=0]{fig3.eps}
\end{minipage}
	\caption{Average background subtracted pulse profiles obtained with (top to bottom) the \textsc{XRT} and \emph{NuSTAR} data for Obs~1 and Obs~2~(left) whereas in the right plot we show pulse profiles for Obs~3 and Obs~4.~Error bars represent 1~${\sigma}$ uncertainties.~For plotting purposes the profiles in different observations were aligned together (refer text for details). }
\label{Avg-Obs12}
\end{figure*}

\begin{figure*}
\centering
\begin{minipage}{0.45\textwidth}
\includegraphics[height=1.\columnwidth,angle=-90]{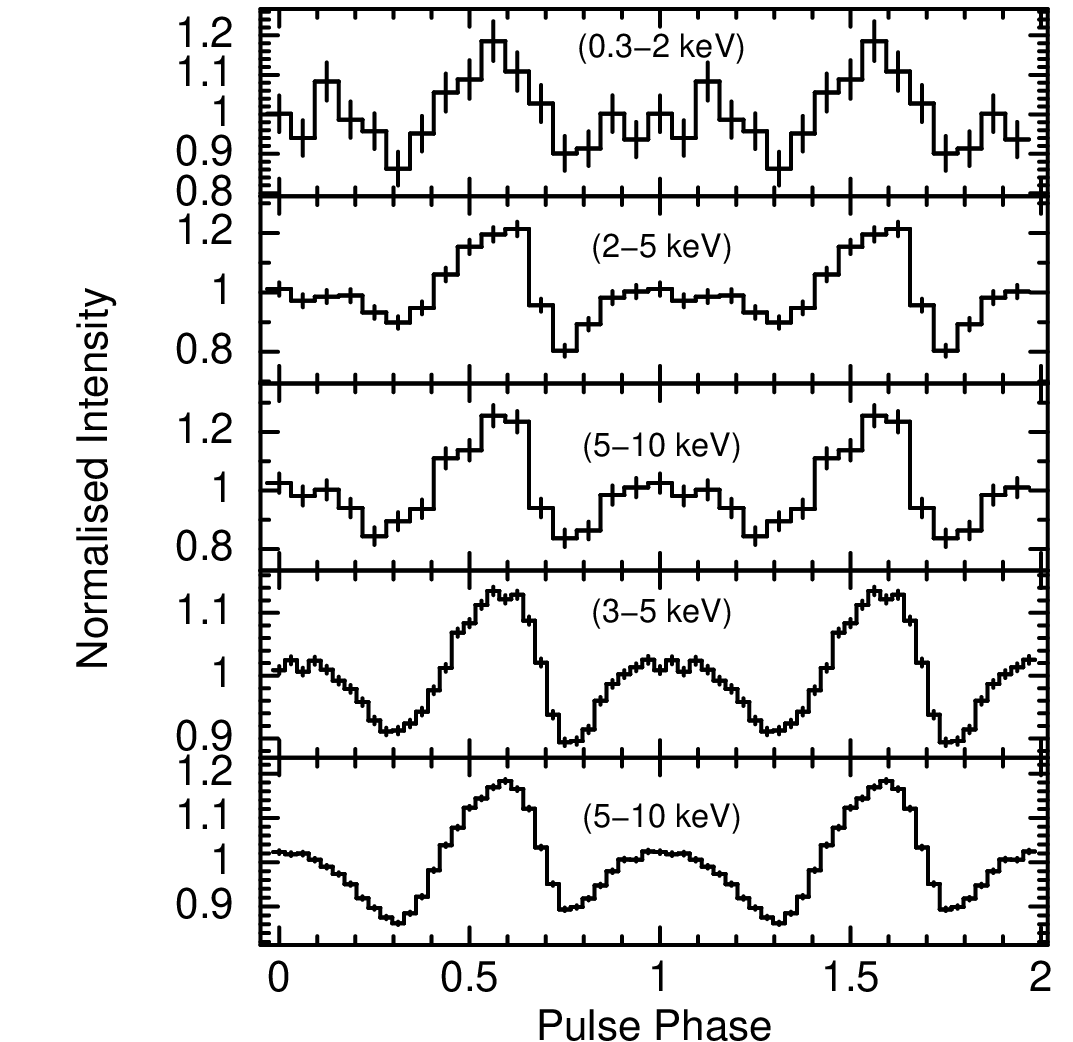}
\end{minipage}
\hspace{0.02\linewidth}
\begin{minipage}{0.45\textwidth}
\includegraphics[height=1.\columnwidth,angle=-90]{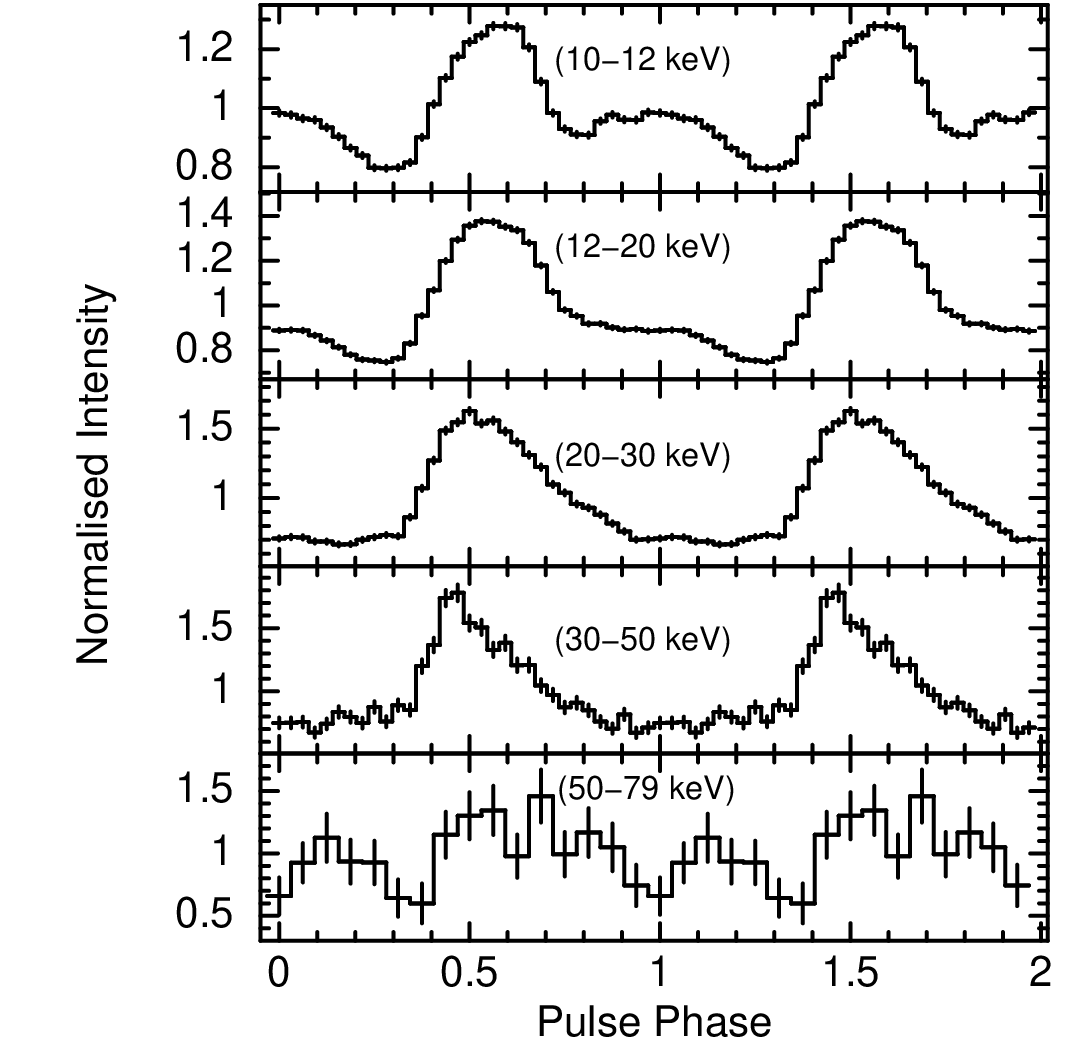}
\end{minipage}
\caption{Energy-resolved pulse profiles of 4U~1901+03 obtained with the \textsc{XRT} (0.3--10 keV) and \emph{NuSTAR} (3--79 keV) data during Obs~1.}
\label{EN-pp1}
\end{figure*}

\begin{figure*}
\centering
\begin{minipage}{0.45\textwidth}
\includegraphics[height=1.\columnwidth,angle=-90]{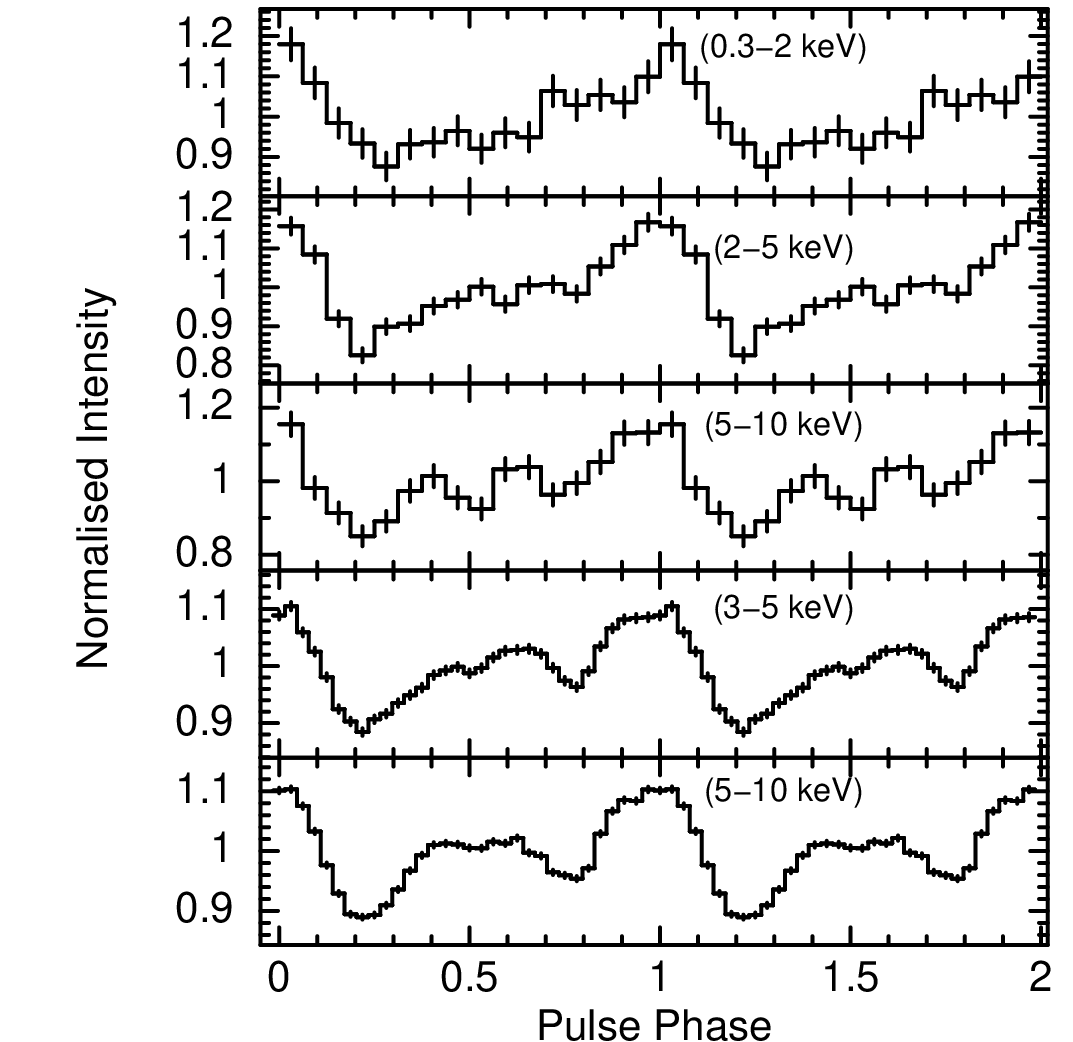}
\end{minipage}
\hspace{0.02\linewidth}
\begin{minipage}{0.45\textwidth}
\includegraphics[height=1.\columnwidth,angle=-90]{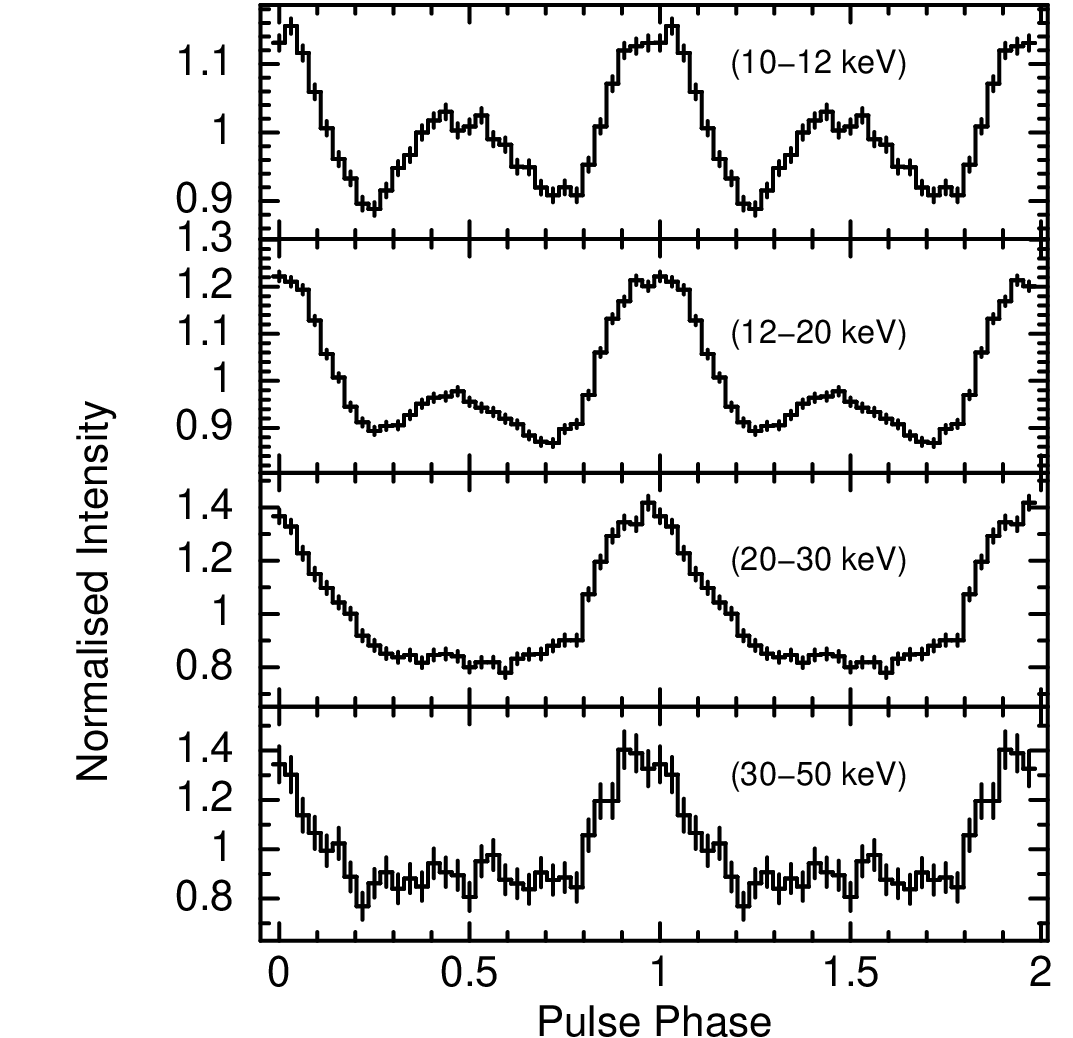}
\end{minipage}
\caption{Energy-resolved pulse profiles of 4U~1901+03 obtained with the \textsc{XRT} (0.3--10 keV) and \emph{NuSTAR} (3--79 keV) data during Obs~2.}
\label{EN-pp2}
\end{figure*}

\begin{figure*}
\centering
\begin{minipage}{0.45\textwidth}
\includegraphics[height=1.\columnwidth,angle=-90]{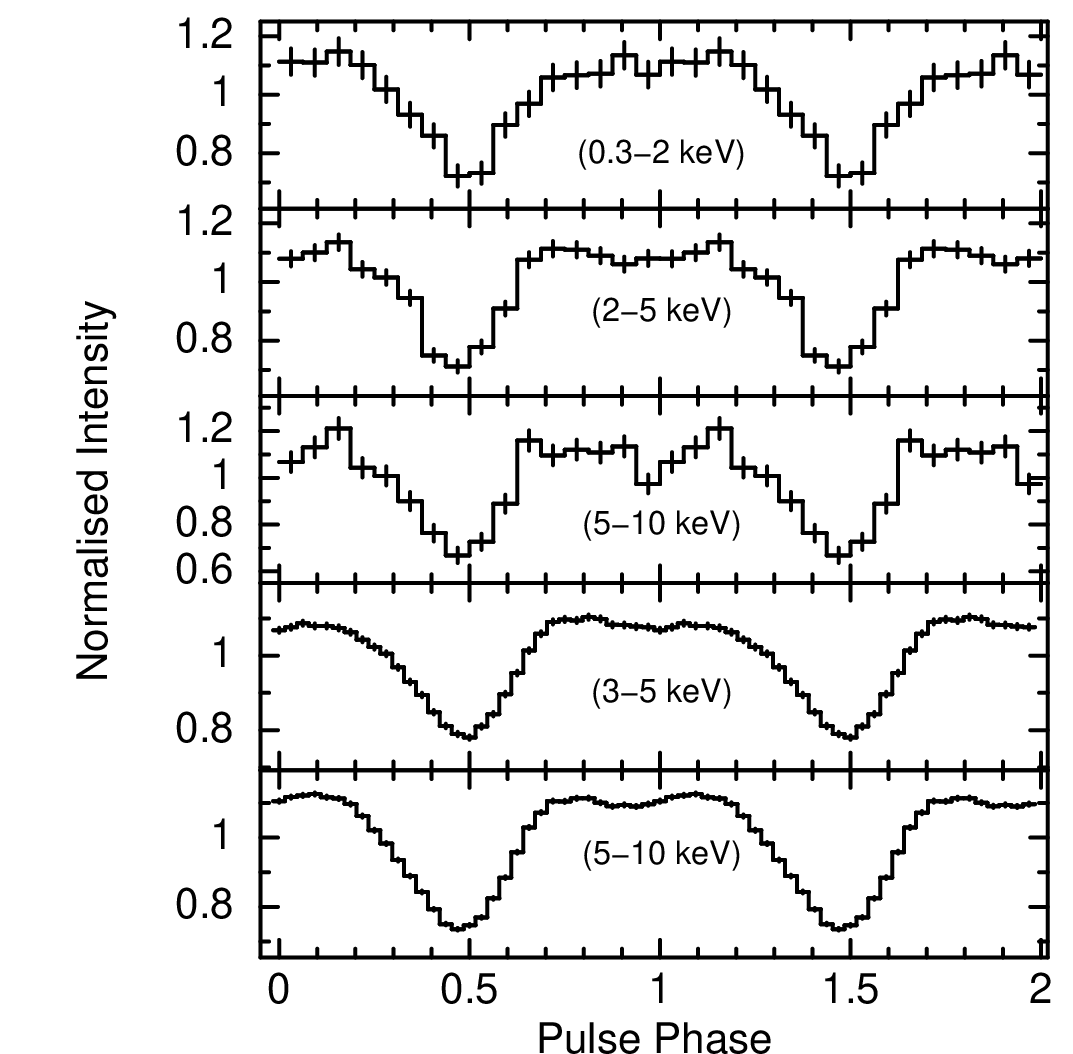}
\end{minipage}
\hspace{0.02\linewidth}
\begin{minipage}{0.45\textwidth}
\includegraphics[height=1.\columnwidth,angle=-90]{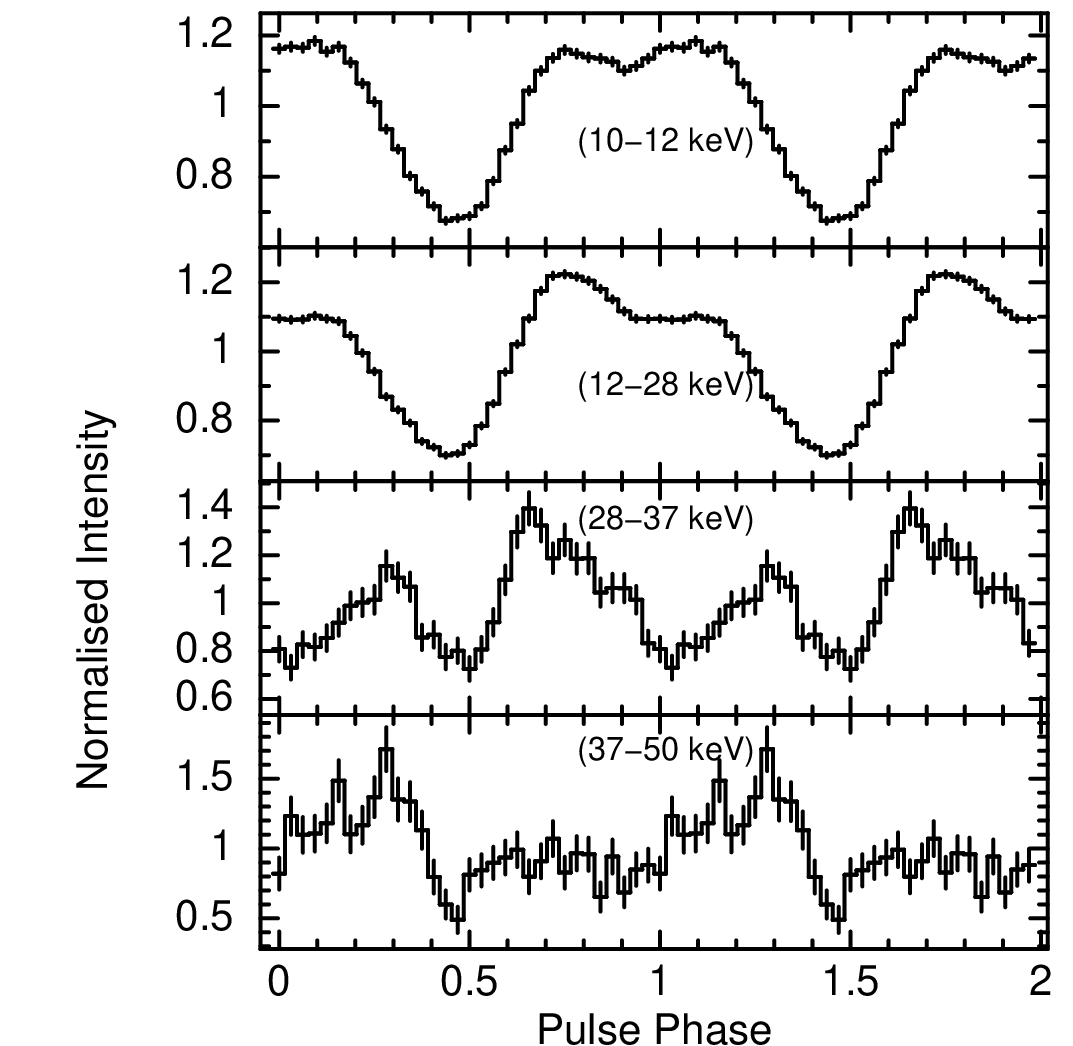}
\end{minipage}
\caption{Energy-resolved pulse profiles of 4U~1901+03 obtained with the \textsc{XRT} (0.3--10 keV) and \emph{NuSTAR} (3--79 keV) data during Obs~3.}
\label{EN-pp3}
\end{figure*}

\begin{figure*}
\centering
\begin{minipage}{0.45\textwidth}
\includegraphics[height=1.\columnwidth,angle=-90]{fig12.eps}
\end{minipage}
\hspace{0.02\linewidth}
\begin{minipage}{0.45\textwidth}
\includegraphics[height=1.\columnwidth,angle=-90]{fig13.eps}
\end{minipage}
\caption{Energy-resolved pulse profiles of 4U~1901+03 obtained with the \textsc{XRT} (0.3--10 keV) and \emph{NuSTAR} (3--79 keV) data during Obs~4.}
\label{EN-pp4}
\end{figure*}

\begin{figure*}
\includegraphics[height=\columnwidth,angle=-90]{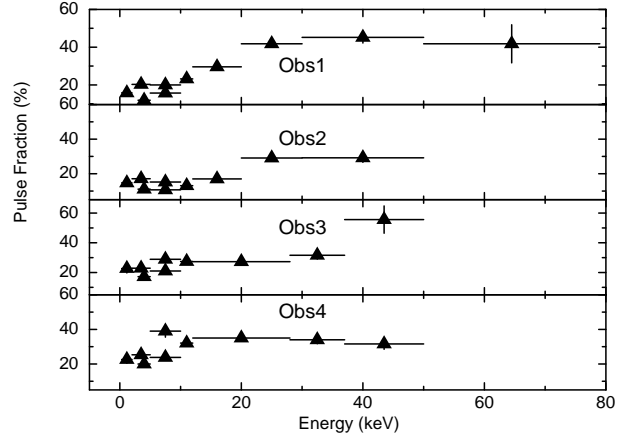}
\caption{This plot shows the variation of pulse fraction as a function of energy during Obs~1,2,3 and 4 (top to bottom).~The horizontal  bars on data points represent the energy ranges in which pulse fraction has been estimated whereas the vertical bars represent errors on the pulse fraction in an energy range. }
\label{PF}
\end{figure*}

\begin{figure*}
\centering
\begin{minipage}{0.45\textwidth}
\includegraphics[height=1.25\columnwidth,angle=0]{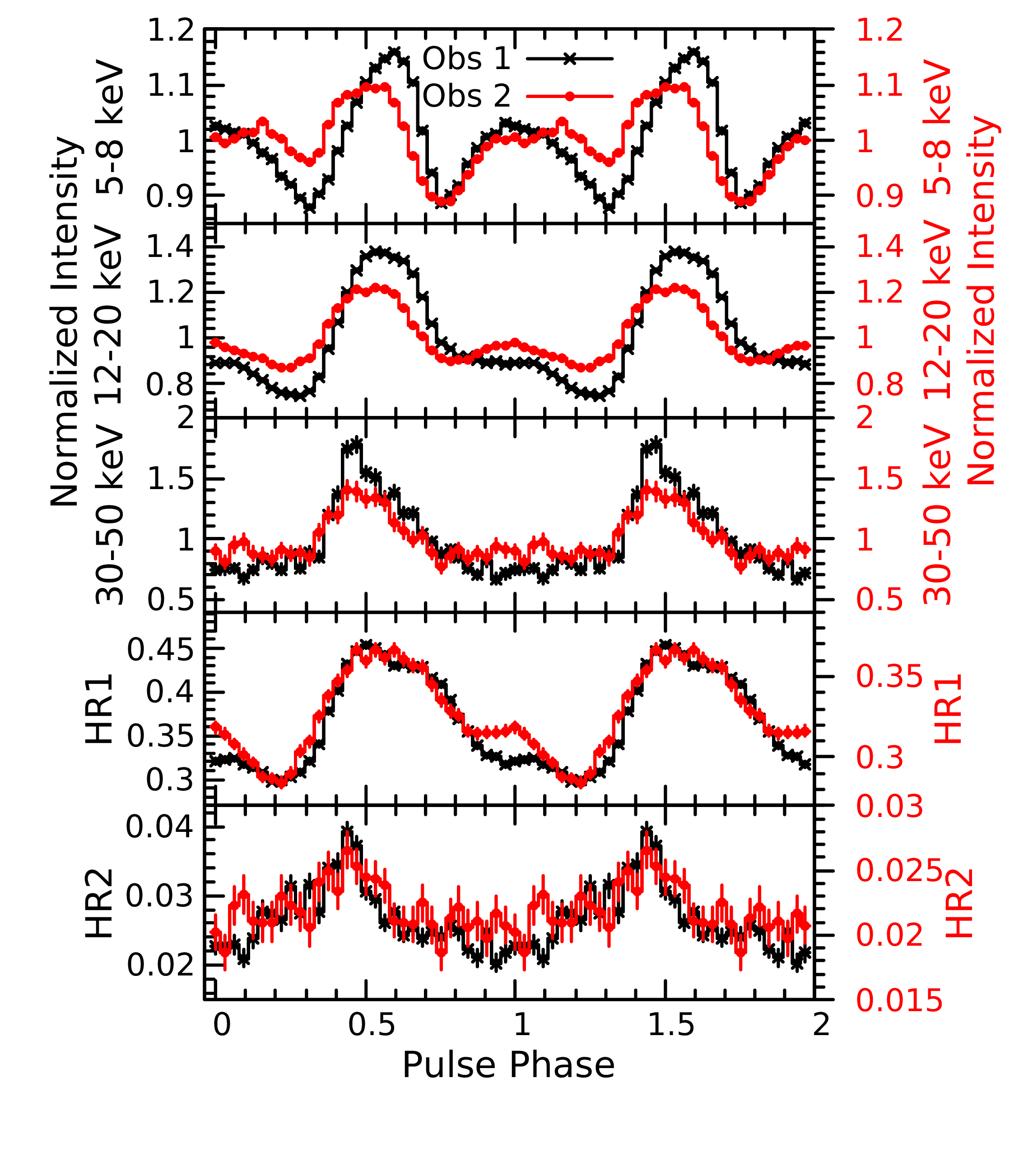}
\end{minipage}
\hspace{0.02\linewidth}
\begin{minipage}{0.45\textwidth}
\includegraphics[height=1.25\columnwidth,angle=0]{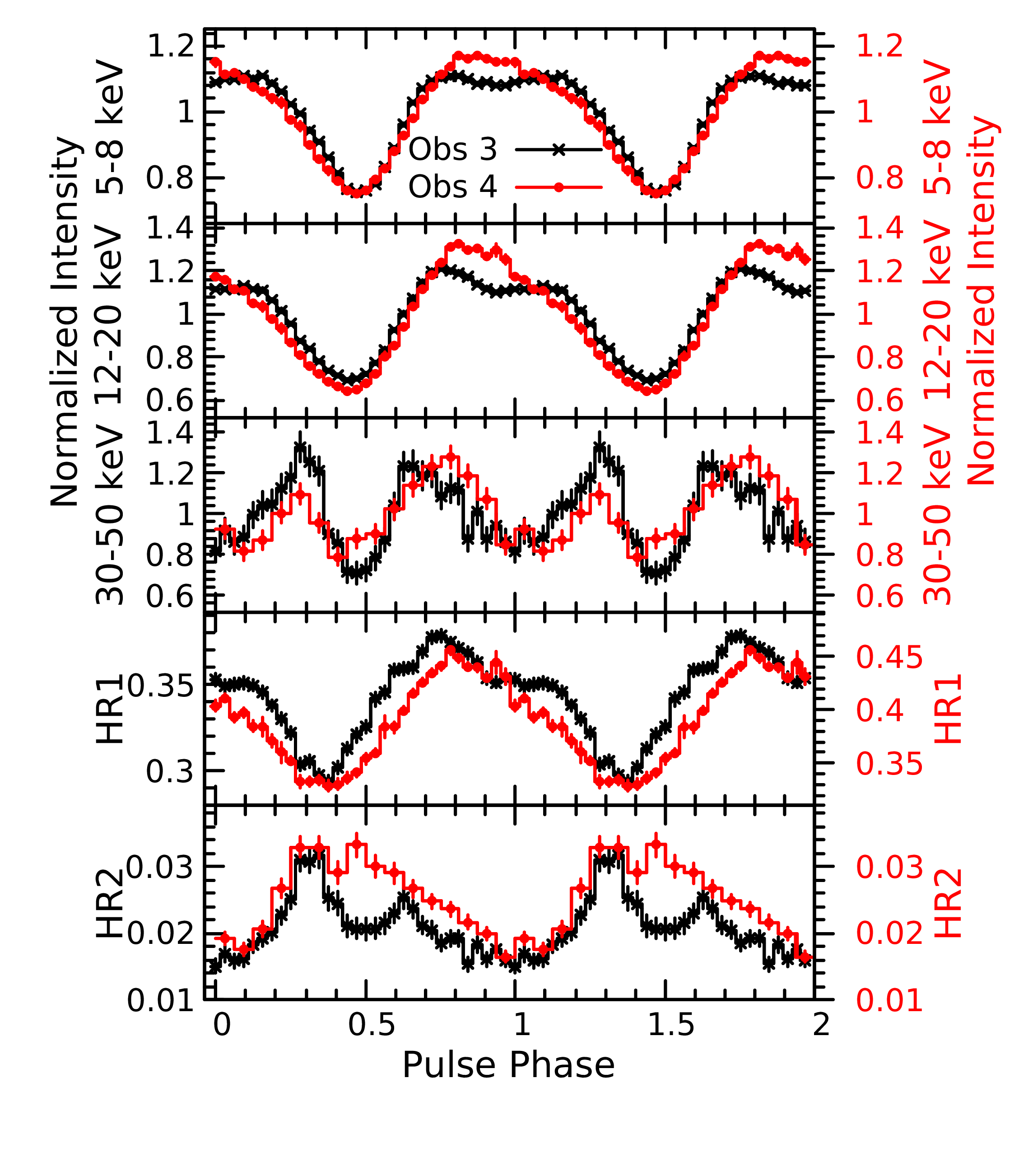}
\end{minipage}
	\caption{In the bottom two panels we show the hardness ratio of energy-resolved pulse profiles of 4U~1901+03 obtained with \emph{NuSTAR} data during Obs~1, 2~(left) and Obs~3,4~(right).~Top three panels of each plot shows energy-resolved pulse profiles used for obtaining these hardness ratios.~In the left figure, Y1 scale~(left) represents data points of Obs~1 while Y2~scale~(right) is for Obs2, similarly, for figure in the right-hand side Y1 scale represents data of Obs3 while Y2 scale is for Obs~4. For plotting purposes the profiles in different observations were aligned together.}
\label{HR1}
\end{figure*}

\begin{figure*}
\centering
\begin{minipage}{0.45\textwidth}
\includegraphics[height=1.25\columnwidth,angle=0]{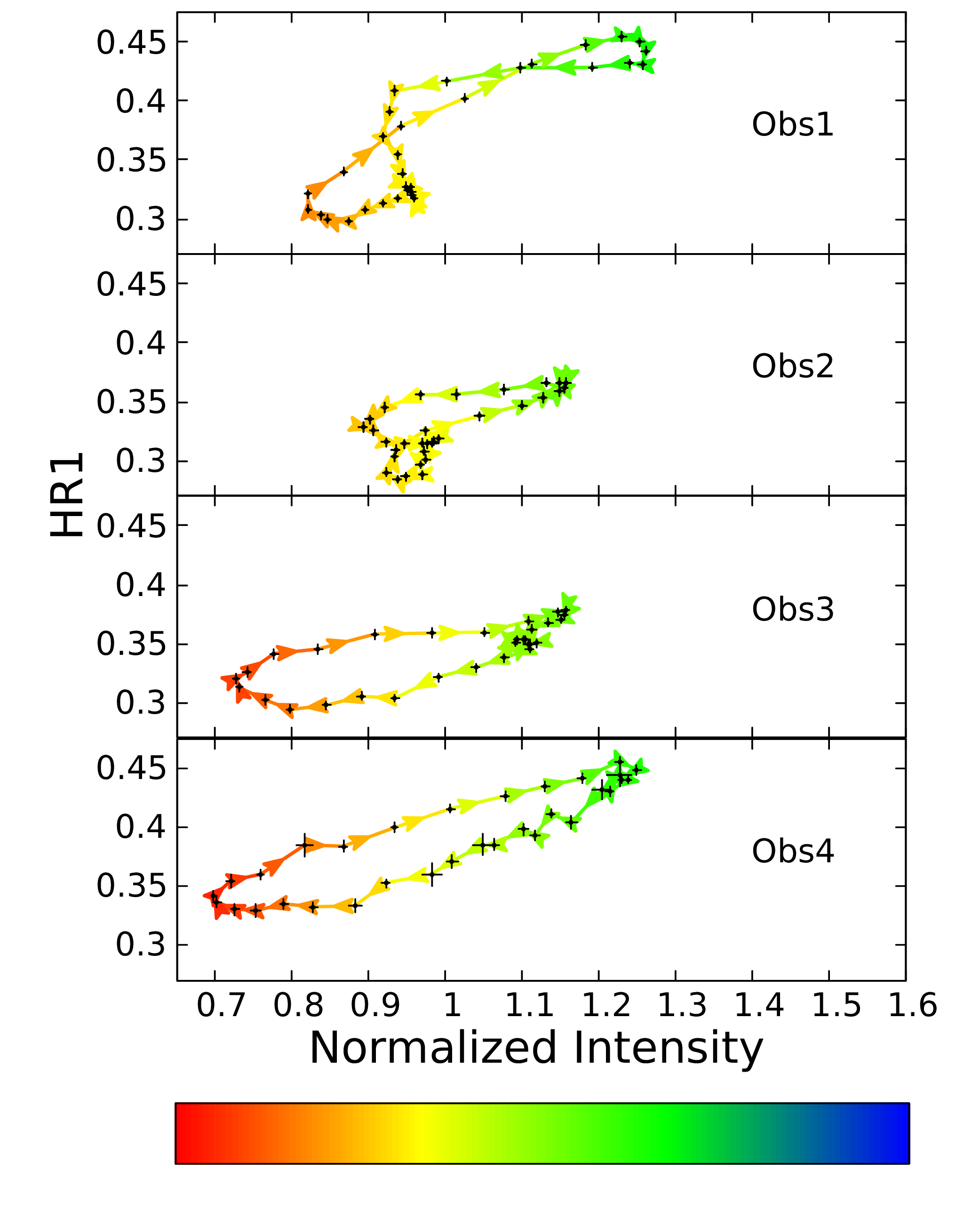}
\end{minipage}
\hspace{0.02\linewidth}
\begin{minipage}{0.45\textwidth}
\includegraphics[height=1.25\columnwidth,angle=0]{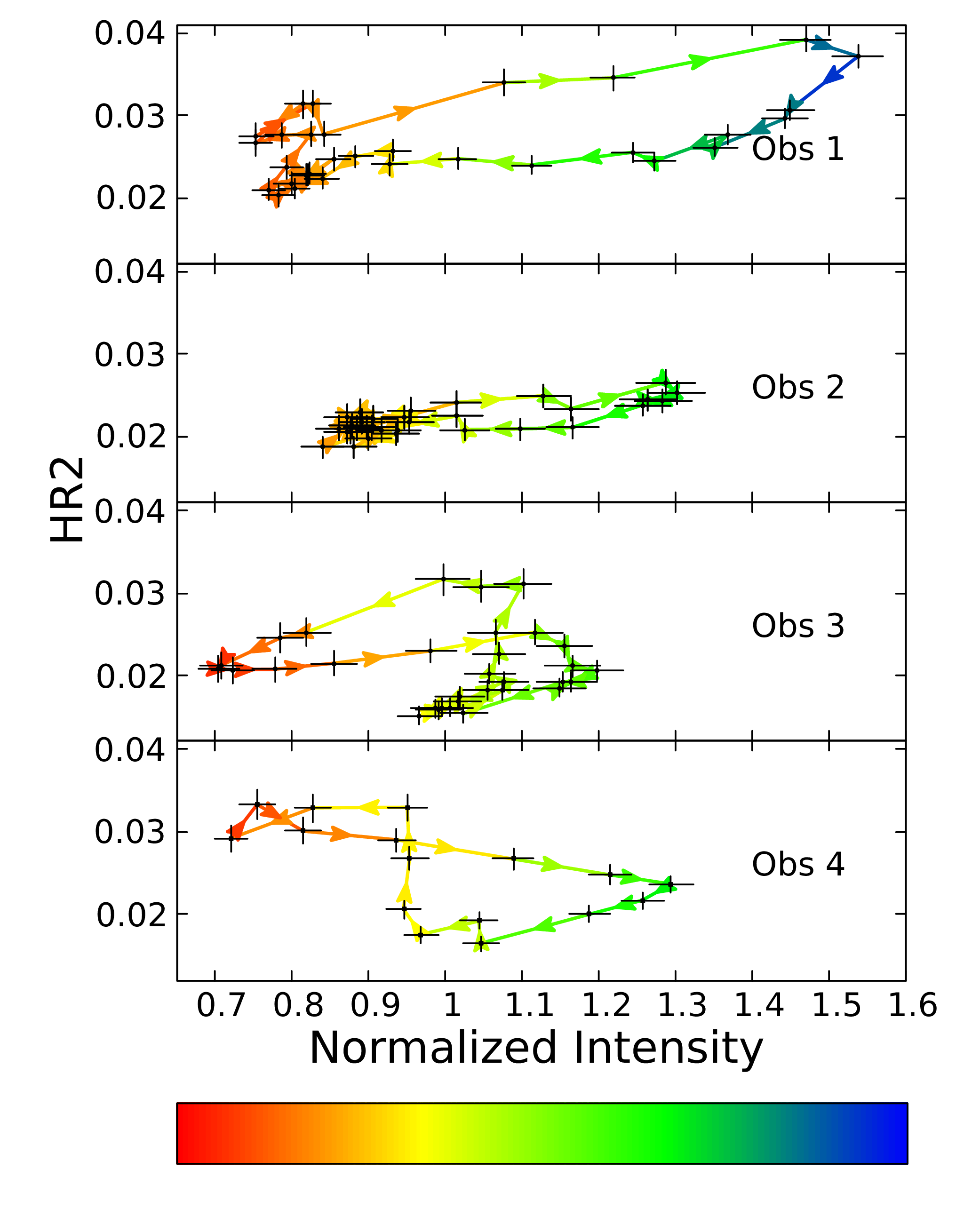}
\end{minipage}
	\caption{In the left we show hardness intensity diagram~(HID) for HR1 whereas in the right plot we show HID for HR2~(see text for details).}
\label{HID-1}
\end{figure*}

\section{Observations and data analysis}\label{sec:dataanalysis}


During the 2019 outburst of 1901, four \emph{NuSTAR} observations were performed, as detailed in Table~\ref{tab:obs}.
Figure~\ref{outburst} shows the Swift--Burst Alert Telescope (BAT) light curve of 1901. We have marked times during which the \emph{NuSTAR} observations were made.
We have also marked \emph{Swift} observations that were contemporaneous to these \emph{NuSTAR} observations. \\

\textsc{HEASOFT} v6.19 and \textsc{NUSTARDAS}~(V1.9.1) was used for standard processing and the extraction. The light curves, spectra, and the response files were extracted using \textsc{NUPRODUCTS}.
A circular region of 100 arc-second radius for the source and a source free circular region of 100 arc-second radius on the same chip for the background files was used.

We have used data obtained with \textsc{XRT} on-board Neil Gehrels \emph{Swift} (see Table~\ref{tab:obs} for details). These \textsc{XRT} observations were performed in the windowed timing (WT) mode due to high source count rate. \textsc{xrtpipeline} was used to reprocess the data and using \textsc{xselect} source and background light curves and spectral files were generated.
In order to avoid possible spectral residuals in the spectra, we have used grade 0 events 
and the source events were obtained from a circular region using of 47{\arcsec} radius.
For the background events the outer source-free regions were used, with a circular region of same radius as that used for the extraction of source events. 

\section{Timing Analysis \& Results}
\subsection{Average Pulse Profiles}

We extracted background and source light curves with a binsize of 100~\rm{ms} in the energy band of 0.3--10~keV and 3--79~keV
using data of \emph{Swift}-\textsc{XRT} and \emph{NuSTAR}, respectively.~Barycentric correction was applied to the background subtracted light curves using {\textit{`barycorr'}}\footnote{https://heasarc.gsfc.nasa.gov/ftools/caldb/help/barycorr.html}, \textsc{FTOOL} task of \textsc{HEASOFT}. We searched for X-ray pulsations using the \textsc{FTOOL} task 
{\textit{`efsearch'}} (see Table~\ref{tab:obs}).~As each observation spanned less than $5{\%}$ of the orbital period, the expected
pulse smearing due to the change in pulse period is negligible.~We performed a correction for smearing due to the orbital motion, using the ephemeris obtained from \emph{FERMI}--\textsc{GBM} and found no change in the shape of pulse profiles. Therefore, we ignored it for rest of the analysis. \\

We  folded \emph{Swift}-\textsc{XRT} and \emph{NuSTAR} light curves into 16 and 32 phase bins, respectively with our measured periods (see Figure~\ref{Avg-Obs12}).
It can be observed that the pulse profiles showed a strong intensity dependence.~During the rising phase of the outburst~(Obs~1) both \emph{Swift}-\textsc{XRT} and \emph{NuSTAR} pulse profiles showed the presence of a primary and secondary peak.~Pulse profiles of both  Obs~1 and Obs~2 have been shifted in pulse phase such that primary
peak appears within 0.3-0.7 phase range.~For Obs~2, the \textsc{XRT} pulse profiles did not show the presence of a secondary peak while this feature exist in the \emph{NuSTAR} profiles, indicating energy dependence of pulse profiles (discussed in detail in the later section). \\

Towards the declining phase of the outburst, pulse profiles tend to be simpler (see right plot of Figure~\ref{Avg-Obs12}).~For Obs~3, both \textsc{XRT} and \emph{NuSTAR} profiles exhibit broad shoulder-like structure while for Obs~4, the profiles tend to be sinusoidal.~Here, again we have shifted all pulse profiles of Obs~3,~4 such that the dip occurs at phase 0.5.

\subsection{Energy-Resolved Pulse Profiles}

To probe into the energy evolution of pulse profiles and pulsar emission 
geometry, we have folded energy-resolved lightcurves in different bands
(see Figures~\ref{EN-pp1},\ref{EN-pp2}, \ref{EN-pp3}, \ref{EN-pp4}). \\

For the rising phase observation~(Obs~1),~there exist a primary and secondary peak in all the profiles below
12~keV, however, it was found that above 12~keV the secondary peak fades away and
pulse profile tend to be a single peaked~(see Figure~\ref{EN-pp1}).~For Obs~2, it was seen that \textsc{XRT} profiles are single peaked in all the energy bands while for the profiles created
with \emph{NuSTAR} there is an evolution of the secondary peak with energy.
At energies below 10~keV, profiles are more consistent with the structures
observed with \emph{Swift}--\textsc{XRT} while 10-12~keV band profile showed a presence 
of a secondary peak which fades away with the increase in energy~(see Figure~\ref{EN-pp2}). \\

For Obs~3~(made during the declining phase of the outburst), we notice that pulse profiles in all the energy bands below 28~$\rm{keV}$ showed a broad shoulder-like structure.~A drastic change was observed in the pulse profiles of 28-37~$\rm{keV}$ and 37-50~$\rm{keV}$ band.~Pulse profile in the 28--37~$\rm{keV}$ band changed from single to twin-peaked.~The first peak appeared at around 0.3 pulse phase while the second peak was seen around 0.7 phase.
In the last energy band~(37--50~$\rm{keV}$), pulse profile again changed to single
peaked profile with a disappearance of the second peak around 0.7 phase~(see Figure~\ref{EN-pp3}).
Similar energy dependence was also seen in the pulse profiles of Obs~4 where all the profiles below 28~$\rm{keV}$ showed a sinusoidal behavior.~We observe that in the 28--37~$\rm{keV}$ band, pulse profile tend to show a sharper peak compared to rest of the energy bands and with some phase shift.~Moreover, there is an emergence of a secondary peak between 0.7-0.8 pulse phase.~It was interesting to notice that in comparison to the 28--37~$\rm{keV}$ band, the pulse profiles of 37-50~$\rm{keV}$ band showed a sharp peak between 0.6-1.0 pulse phase with a disappearance of twin-peaked behavior~(see Figure~\ref{EN-pp4}). \\

Pulse fraction~(PF) defined as the ratio between
the difference of maximum ($I_{max}$ ) and minimum ($I_{min}$ ) intensity to their sum was estimated: $((I_{max} - I_{min})/(I_{max} + I_{min}))$.~We investigated the change in
pulse fraction with energy for which we estimated PF in several energy ranges~(see Figure~\ref{PF}).
The horizontal bars on data points in Figure~\ref{PF} represent the energy ranges in
which pulse fraction has been estimated while the vertical bars represent errors on the pulse fraction in a given energy range.~We found that for all observations PF shows an increasing trend with energies up to 25~{\rm{keV}}.~For Obs~1, PF increased from $11.9\pm0.1$ to $45.0\pm3.0$ $\%$, for Obs~2, it increased from $11.1\pm0.1$ to $29.0\pm2.0$ $\%$.~PF showed values between $17.2\pm0.2$ and $55.0\pm9.0$ $\%$ during Obs~3 while for Obs~4 there was an increase from $20.0\pm1.0$ to $34.0\pm2.0$ $\%$ in the PF.

\subsection{Hardness Ratio}
To track spectral evolution with the pulse phase, we have used a model independent way of creating hardness ratio~(HR) plots.~Figure~\ref{HR1} shows phase-resolved HR.~We computed two HR indices,~HR1 and HR2.~HR1 was calculated by taking ratio of count rates in 12--20~$\rm{keV}$ and 5--8~$\rm{keV}$ energy bands while HR2 was computed for count rate ratio between 30--50~$\rm{keV}$ and 12--20~$\rm{keV}$ energy band.~HR was computed across 32 phase bins except for Obs~4 where HR2 was computed using only 16 phase bins (due to the limited statistics). \\
We notice that HR plots for Obs~1 and 2 are similar for both indices showing a single peaked profile, where the peak corresponds to the hard spectra.~Comparing HR plots to the pulse profiles, it appears that peak of the HR plot corresponds to peak of the pulse profiles.
For the case of Obs~3 and 4, it is interesting to note that the peak of HR1 corresponds to the dip of HR2.~The most intriguing aspect observed in Obs~3 is that HR2 profiles showed a twin-peaked behaviour,~indicating significant changes in the hardness ratio or spectral behavior of two HR indices, HR1 and HR2.  
Figure~\ref{HID-1} shows Hardness-Intensity diagrams obtained by plotting phase-resolved hardness ratio against normalized sum intensity in different energy bands.~We
observed a direct correlation between intensity and hardness in all except HR2 of Obs~4, suggesting an increase in hardness with the increase in intensity.  \\

\section{Spectral Analysis}

\subsection{Phase-Averaged Spectral Analysis}

We performed X-ray spectral analysis using the combined data of \emph{Swift}--\textsc{XRT} and \emph{NuSTAR}, during all the four observations~(Obs~1,~2,~3~\& 4). The X-ray spectral fitting package, \textit{XSPEC}~\citep[version--12.9.0;][]{Arnaud1996} was used.~The extracted spectra were grouped to have a minimum of 25 counts per bin using the \textsc{FTOOL} task `\textit{grppha}'. We have used the following energy ranges for spectral fitting:~1--10~keV and 3--70~keV of \textsc{XRT} and \textsc{NuSTAR}, respectively.~The \textsc{XRT} data below 1~keV was ignored due to spectral residuals below in the WT mode spectra \footnote{http://www.swift.ac.uk/analysis/xrt/digest{$\_$}cal.php${\#}$abs}.~To include the effects of Galactic absorption we have used the model component `\textit{tbabs}' with abundances from \citet{Wilms2000} and crossections as given by \citet{Verner1996}. A multiplicative term~(\textit{CONSTANT}) 
was also added to the model to account for calibration uncertainities
between the \textsc{XRT} and \emph{NuSTAR}-\textsc{FPMA} and \textsc{FPMB}. The value was fixed to 1 for the \emph{NuSTAR}-\textsc{FPMA} and was allowed to vary for \textsc{FPMB}
and \textsc{XRT}. All fluxes were estimated using the convolution model `\textit{CFLUX}'. \\

\begin{figure*}
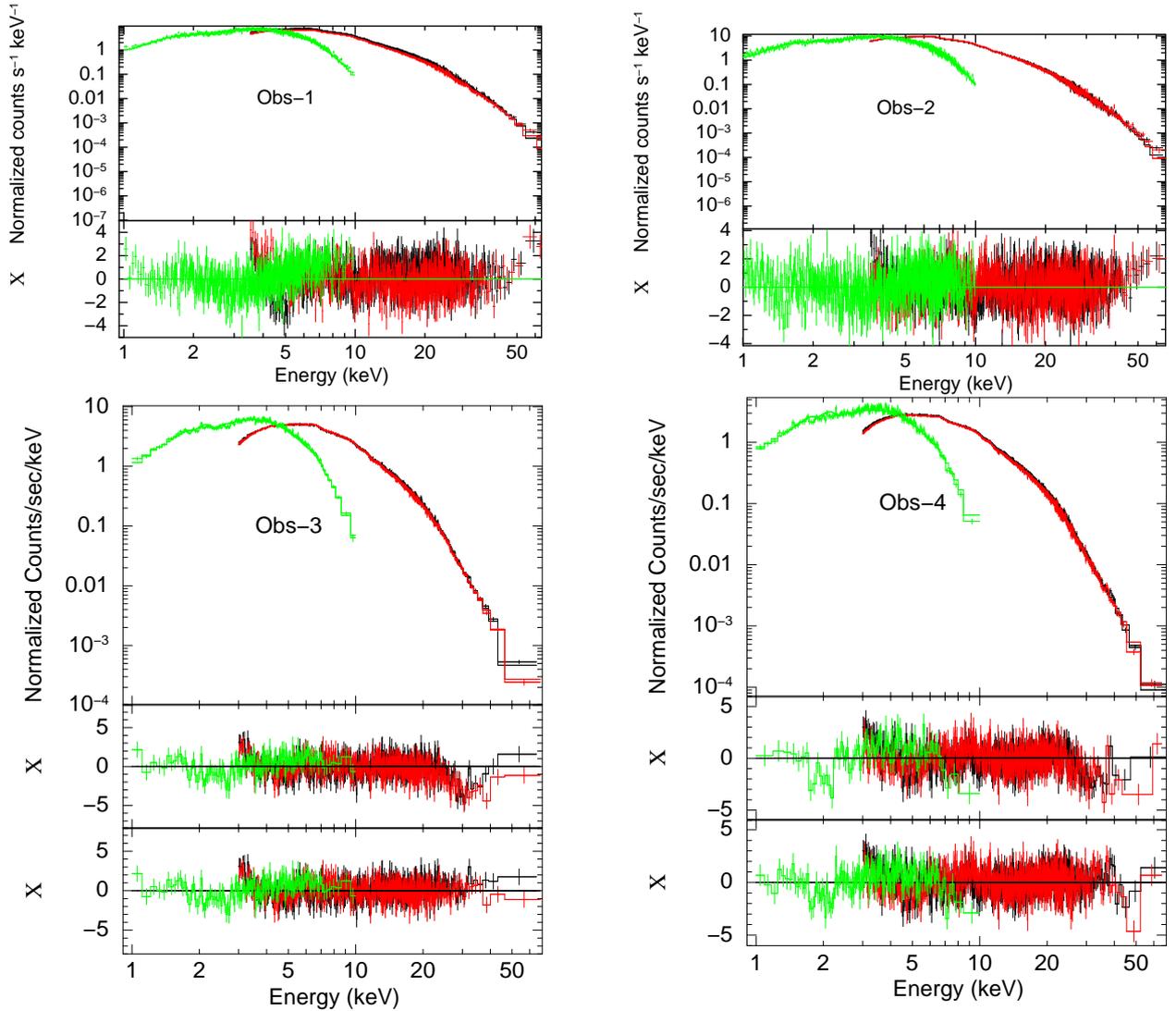

\centering
\begin{minipage}{0.45\textwidth}
\includegraphics[height=1.\columnwidth,angle=-90]{fig27.eps}
\includegraphics[height=1\columnwidth,angle=-90]{fig28.eps}
\end{minipage}
\hspace{0.04\linewidth}
\begin{minipage}{0.45\textwidth}
\includegraphics[height=1\columnwidth, angle=-90]{fig29.eps}
\includegraphics[height=1\columnwidth,angle=-90]{fig30.eps}
\end{minipage}
\caption{Phase-averaged spectra obtained with \emph{Swift}-\textsc{XRT} and \emph{NuSTAR} during four observations made during 2019 outburst of 1901. The residuals in the middle panel of bottom two plots indicates the presence of a Cyclotron resonance scattering feature~(CRSF).~The plot has been created using Model~2 spectral fit parameters.}
\label{spec}
\end{figure*}

X-ray spectra of X-ray binary pulsars are thermal in nature formed
in a hot plasma~(T$\sim$ $10^8$ $\rm{K}$) over the magnetic poles of the neutron star. The emission process is governed by the Comptonization of thermal photons which gain energy by scattering off hot plasma electrons. Typically, shape of the X-ray spectral continuum is described by phenomenological power law models with an exponential cut-off at higher energies \citep[see e.g.,][]{Coburn2002}. There exist several \textsc{XSPEC} models
such as a cut-off power law (`\textit{CUTOFFPL}'), high energy cut-off power law (`\textit{HIGHECUT}'), a combination of two negative and positive power laws with exponential cut-off (`\textit{NPEX}'). Other local models like power law with Fermi-Dirac cut-off \citep[`\textit{FDCUT}'][]{Tanaka1986} and a smooth high energy cut-off model \citep[`\textit{NEWHCUT}'][]{Burderi2000}.
We have tried all these traditional models (mentioned above), in addition to the thermal Comptonization model `\textit{COMPTT}' \citep{Titarchuk94} to fit the continuum X-ray emission of 1901 during these four observations.~Adding only continuum components to the X-ray spectra of all four observations showed the presence of a deep, broad negative residual around 10~{\rm{keV}}.~The ``10~\rm{keV} feature" in the \emph{NuSTAR} spectra was also reported by \citet{Coley19} during the 2019 outburst of 1901.~Similar feature was also seen in the \emph{RXTE} observations made during the 2003 outburst of 1901 and was interpreted as a possible cyclotron line in this source \citep[see][for details]{Reig16}.~Therefore, we added \textsc{XSPEC} model \textit{GABS}
which has three parameters: the line energy~($E$), width~(${\sigma}$) and a  normalisation  coefficient~(norm).~The normalisation corresponds to the line depth and is related to the optical depth, which at the line center is given by ${\tau}=norm/({\sqrt{2}{\pi}{\sigma}})$.
The line was centered around 10~$\rm{keV}$~($E_{Gabs\_add}$).~Adding this model component improves the fit.~Furthermore, spectral residuals indicated the presence of emission features between 6 and 7 ${\rm keV}$.~Iron  line  emission  is  commonly observed in the X-ray spectra of accreting X-ray pulsars \\
\citep[see e.g.,][]{Basko1978,Basko1980,Ebisawa1996}.~X-ray photons emitted from the pulsar interact with the ionized/neutral iron atoms emitting characteristic emission  lines.~Therefore, we added a Gaussian component centered at around 6.5~$\rm{keV}$~($Fe~K_{\alpha}$).~The iron line energy~($E_{Fe}$) did not show any significant change from Obs~1 to Obs~4.
However, the equivalent width~(EW) of this line was found to be variable during these observations.~The maximum value of EW was observed during Obs~2.~A blackbody component was also needed to fit the low energy excess seen in all four observations.~An interesting observation was the presence of negative residuals around 30~$\rm{keV}$~($E_{cyc}$) in the spectral residuals of Obs~3 and 4 only (see Figure~\ref{spec}).~Therefore, we added an another `\textit{GABS}' component to the X-ray spectra of these two observations (see Table~\ref{spec-para34} for best-fit values).~Addition of this component reduced the value of ${\chi_{\nu}}^{2}$ from 1.20 to 1.14 (2064 dof) for Obs~3 and 1.20 to 1.14 (2064 dof) for Obs~4.~The best-fit values obtained from
the spectral fitting are given in Tables~\ref{spec-para12} \& \ref{spec-para34}. \\

It is often observed that spectra of X-ray pulsars with high photon statistics
show deviations from simple phenomenological models and a wave-like feature in the fit residuals is present between 10--20~$\rm{keV}$~(commonly known as ``$10 \rm{keV}$ feature").~The origin of this feature is still under discussion and remains unclear \citep[see e.g.,][]{Coburn2002,Staubert2019,Bissingerne2020}.
Therefore, we attribute $30~{\rm keV}$ feature to CRSF formed by resonant scattering of X-ray photons with electrons quantized in discrete Landau levels. \\

\textit{Statistical significance}: To estimate significance of the detection of the CRSF at around $30~{\rm keV}$, we tried to fit the combined spectra of \emph{Swift} and \emph{NuSTAR} of Obs~3 \& 4 using the following model~(Model~2):   \\
const*tbabs*(cutoffpl*gabs+bbodyrad+gauss+gauss), keeping its power-law index frozen to the value obtained from the best-fitting broadband spectrum.~Addition of the CRSF improved the ${\chi}^{2}$ of Obs~3 from 2439~(2067 dof) to 2392 for 2064~dof and for Obs~4, it changed from 2485~(2066~dof) to 2418~(2064~dof). \\

A test commonly used to infer whether the residuals with and without a model fit is systematic in nature is run-test (also called the Wald-Wolfowitz test) and is often performed to evaluate the statistical significance of a weak absorption feature against the continuum \citep[see e.g.,][]{Orlandini12, Maitra17a}.
To further verify the statistical significance of this absorption
feature around 30~$\rm{keV}$,~we performed run-test~\citep{Barlow89} to derive the null hypothesis of the randomness in the
residuals of spectral fitting in the 20--40~$\rm{keV}$ energy range.
The number of data points used for the run test in the
20--40~$\rm{keV}$ energy range is 17 (10 points below zero and 7 points
above zero) for Obs~3, 22 (16 points below zero and 6 points above zero) for Obs~4.
This gave the probability of obtaining ${N_r} {\leq} 3$ (where, $N_r$ is the number of runs) equals
$0.06{\%}$ and $0.01{\%}$ for Obs~3 \& 4, respectively.~These obtained values
strongly indicate that the residuals are not due to random fluctuations but
have a systematic structure and addition of CRSF component is statistically
significant. \\

It can be seen from Tables~\ref{spec-para12} \& \ref{spec-para34} that all models~(Model~1 to Model~6) provided acceptable fits to the data but Model~2 allowed us to constrain spectral-fit parameters of all the phase-resolved spectra of four observations.~Therefore, we discuss in detail the results obtained with Model~2.~Please note that an additional `\textit{GABS}' component was added to all models while fitting the X-ray spectra obtained with Obs~3 \& 4. \\

The blackbody temperature measured with observations close to the peak of 
outburst~(Obs~1 \& Obs~2) was $\sim$~0.15~$\rm{keV}$ and it decreased to $\sim$ 0.1~$\rm{keV}$ close to the declining phase of the outburst~(Obs~3 \& Obs~4).
We found that photon index showed similar values~(within errorbars) during Obs~1, Obs~2 and Obs~3.~However, a higher value~($\sim$~0.68; Model~2) was observed during Obs~4.
The values of $E_{Gabs\_add}$ lie within the range of $11.4-8.6~\rm{keV}$.~$E_{cyc}$ remained almost constant (within error bars) during Obs~3 and Obs~4. 

Confidence contours were plotted to check the interdependence and to look for possible degeneracies between some of the model parameters of Obs~3.~We used this observation because an additional absorption feature around $30~\rm{keV}$ was significantly found in this and all spectral fit parameters were well constrained compared to Obs~4.~Figure~\ref{contour} shows the ${\chi}^2$ confidence contours between pairs of 
some of the model parameters. It can be seen that photon index (${\Gamma}$) and cut-off energy~($E_{cut}$) are correlated and difficult to constrain independently while $E_{Gabs\_add}$ at 10~keV and cut-off energy~($E_{cut}$) could be well constrained. \\

\subsection{Phase-resolved Spectroscopy}
Motivated by the change in HR as indicated by Figure~\ref{HR1},~we performed a spin-phase resolved spectroscopy to understand
the accretion geometry and surrounding of the pulsar.
We accumulated the phase-sliced spectra in 10 phase bins
using the \textsc{XSELECT} package.~It is worth mentioning that while performing phase-resolved spectroscopy we 
have only used the \emph{NuSTAR}~(\textsc{FPMA} and \textsc{FPMB})~data 
as \emph{Swift} observations were
not strictly simultaneous.~We have used the same response matrices and effective
area files as used for phase-averaged spectroscopy. Spectral studies
were carried out in the 3-70~$\rm{keV}$ energy range using the same technique as opted while
performing phase-averaged spectroscopy. For all four observations, we fixed iron line widths and neutral hydrogen column density to the best fit values obtained for the
phase-averaged spectroscopy.~We observed that for Obs~1, the addition of a blackbody component 
was needed to obtain the best fit while it was not included for rest of the observations.~In the case of obs~3~$\&$~4, we also fixed width~(${\sigma}_{cyc}$) of the $E_{cyc}$ to the best fit phase-averaged values as it was difficult to constrain these while performing phase-resolved spectroscopy. \\

\begin{figure*}
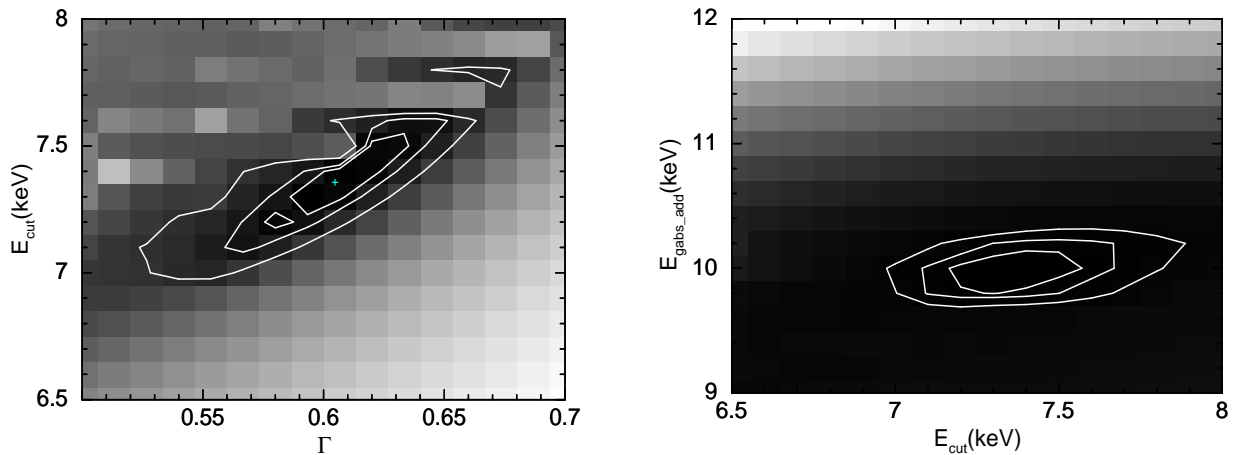

\begin{minipage}{0.45\textwidth}
\includegraphics[height=\columnwidth,width=0.75\columnwidth,angle=-90]{fig31.eps}
\end{minipage}
\hspace{0.02 \linewidth}
\begin{minipage}{0.45\textwidth}
\includegraphics[height=\columnwidth,width=0.75\columnwidth,angle=-90]{fig32.eps}
\end{minipage}
\caption{${\chi}^2$ confidence contours between photon index~($\Gamma$) and cut-off energy~($E_{cut}$) on the left while on the right we show confidence contours between 
$E_{cut}$ and $E_{Gabs\_add}$ at 10~keV, obtained from the phase-averaged spectra of Obs~3. The innermost to outermost contours represent respectively 68${\%}$, 90${\%}$ and 99${\%}$ confidence levels. }
\label{contour}
\end{figure*}

\begin{figure*}
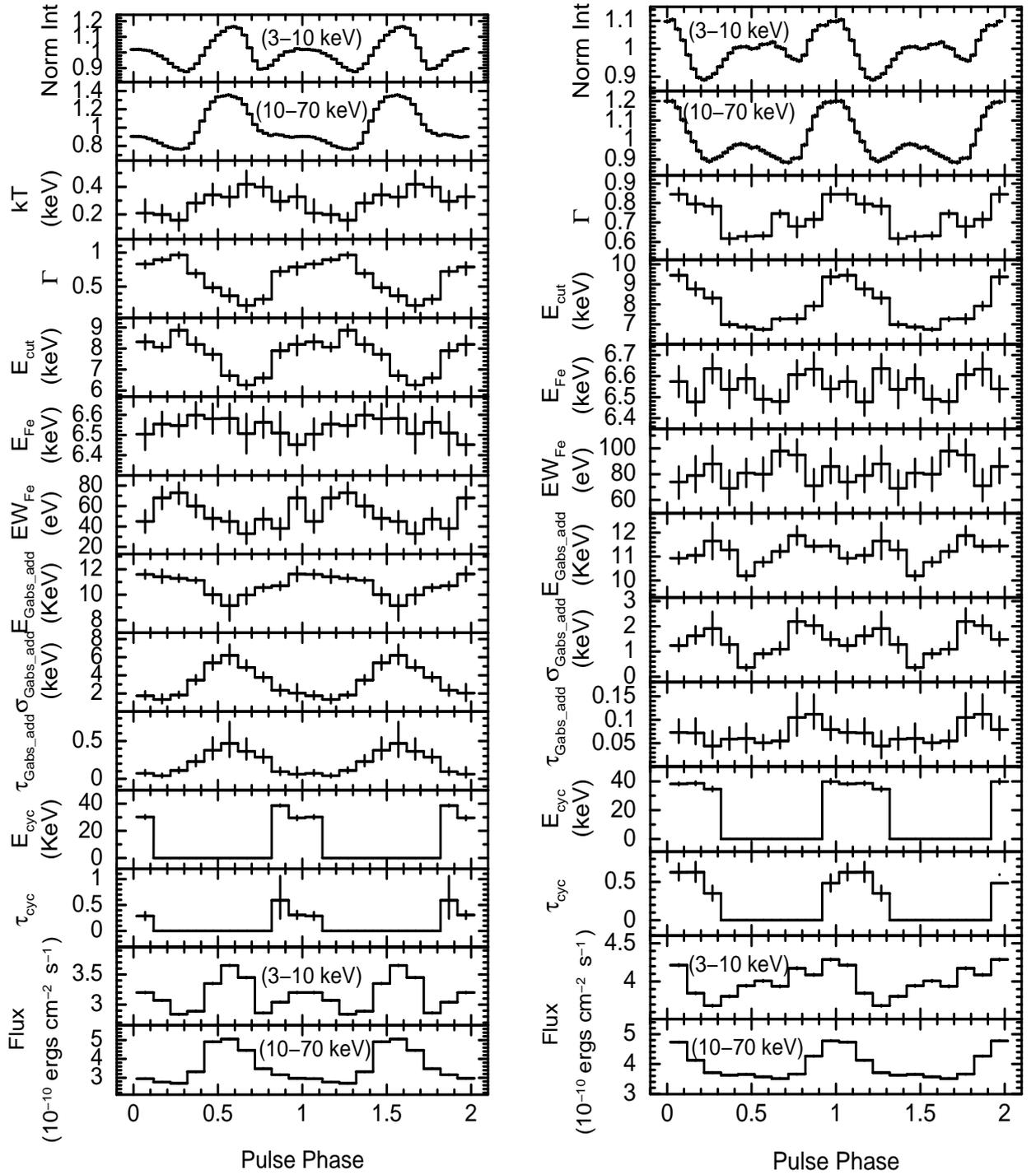

\begin{minipage}{0.45\textwidth}
\includegraphics[height=2.5\columnwidth,width=1.1\columnwidth,angle=0]{fig33.eps}
\end{minipage}
\hspace{0.02 \linewidth}
\begin{minipage}{0.45\textwidth}
\includegraphics[height=2.5\columnwidth,width=1.1\columnwidth,angle=0]{fig34.eps}
\end{minipage}
	\caption{This plot shows variation of spectral parameters across the pulse phase for Obs~1(left) \& 2(right).~We have used Model~2 for performing phase-resolved spectroscopy~(please see text for details).~Here, we would like to mention that the value 0 in $E_{cyc}$ and ${\tau}_{cyc}$ panels corresponds to observations which did not require this component for obtaining the best fit. }
\label{Phase-resol-1-2}
\end{figure*}

\begin{figure*}
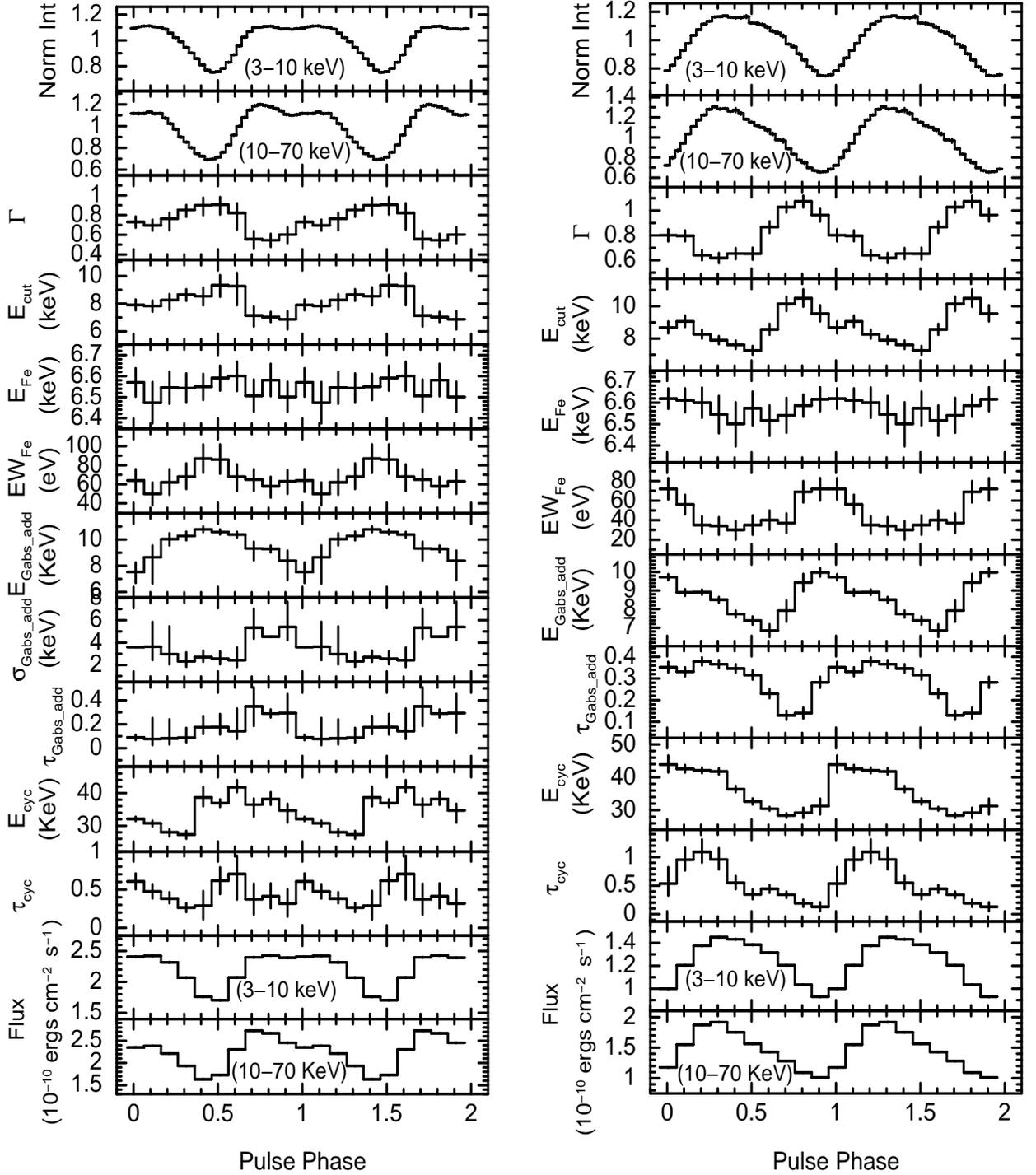

\begin{minipage}{0.45\textwidth}
\includegraphics[height=2.5\columnwidth,width=1.1\columnwidth,angle=0]{fig35.eps}
\end{minipage}
\hspace{0.02 \linewidth}
\begin{minipage}{0.45\textwidth}
\includegraphics[height=2.5\columnwidth,width=1.1\columnwidth,angle=0]{fig36.eps}
\end{minipage}
	\caption{This plot shows variation of spectral parameters across the pulse phase for Obs~3(left) \& 4(right).~Model~2 was used for performing phase-resolved spectroscopy.}
\label{Phase-resol-3-4}
\end{figure*}

\begin{figure*}
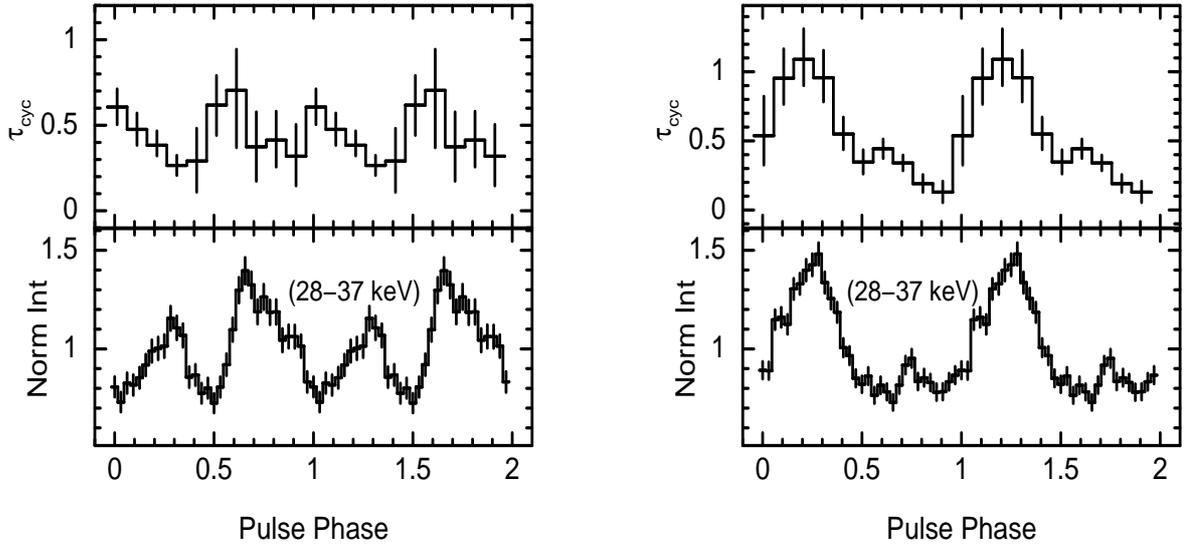

\begin{minipage}{0.45\textwidth}
\includegraphics[height=\columnwidth,width=\columnwidth,angle=0]{fig37.eps}
\end{minipage}
\hspace{0.02 \linewidth}
\begin{minipage}{0.45\textwidth}
\includegraphics[height=\columnwidth,width=\columnwidth,angle=0]{fig38.eps}
\end{minipage}
\caption{This plot shows a similarity between the optical depth~($\tau{cyc}$) and pulse profile in the 28--37~$\rm{keV}$ energy band for Obs~3(left) \& 4(right).}
\label{tau-pp}
\end{figure*}

Figures~\ref{Phase-resol-1-2} \& \ref{Phase-resol-3-4} show spectral parameters 
obtained from the phase-resolved spectroscopy. The top two panels in these plots
show the pulse profiles in 3--10~$\rm{keV}$ and 10--70~$\rm{keV}$ energy bands.
The bottom two panels show corresponding values of the source flux.~In Figure~\ref{Phase-resol-1-2}~(left),~continuum spectral
parameters such as blackbody temperature~($kT$), photon index~($\Gamma$), and high energy cut~($E_{cut}$) are shown in the third, fourth and fifth panel, respectively.~As mentioned earlier, blackbody component was not required for Obs~2,~Obs~3, and Obs~4, therefore, we show $\Gamma$, and $E_{cut}$ in the third, fourth panel, respectively for these observations (see right plot of Figure~\ref{Phase-resol-1-2} and \ref{Phase-resol-3-4}).
It can be clearly seen that all these parameters significantly vary over pulse-phase.~$\Gamma$
and $E_{cut}$ were found to be strongly correlated and might indicate degeneracy between the two model parameters. This is consistent with that observed from the confidence contours between these two parameters (Figure~\ref{contour}).
The blackbody temperature showed a variation between 0.1 and 0.5~$\rm{keV}$
and an anti-correlation was observed between photon index and blackbody temperature for Obs~1.~Photon index varied between 0.2 and 1 during Obs~1 while it showed values between 0.6 and 0.85 during Obs~2.~During all observations, we observed a low value of $\Gamma$ during peak of the pulse profiles, indicating spectra is harder during these phases. This is well consistent with the previous reports of this source and also typical of X-ray pulsars.~Iron line energy~($E_{Fe}$) did not show any variation across the pulse phase, however, we observe some anti-correlation between its equivalent width~($EW_{Fe}$) and pulse profiles.~A clear correlation could be seen between $E_{cut}$ and $E_{Gabs\_add}$ in all observations.
``10~$\rm{keV}$ feature" showed variation across pulse phase in all four observations.~For Obs~1,~2,\&~3,~line centroid energy varied between 9 to 11~$\rm{keV}$ while for Obs~4 it showed a value between 7 and 10~$\rm{keV}$.~There also exist some correlation between ${\tau}_{Gabs_add}$ and pulse profiles.~$E_{cyc}$ varies by ${\sim}~60{\%}$~(22--42$\rm{keV}$) \& $\sim$~47$\%$~(29--44~$\rm{keV}$) for Obs~3 \& 4, respectively and there exist a correlation between pulse profiles (shown in top panels of Figure~\ref{Phase-resol-3-4}), and $E_{cyc}$.~Moreover,~${\tau}_{cyc}$ also showed variation across the pulse phase with a trend similar to that observed in the pulse profiles of 28--37~$\rm{keV}$ band (see Figure~\ref{tau-pp}). \\

\begin{figure*}
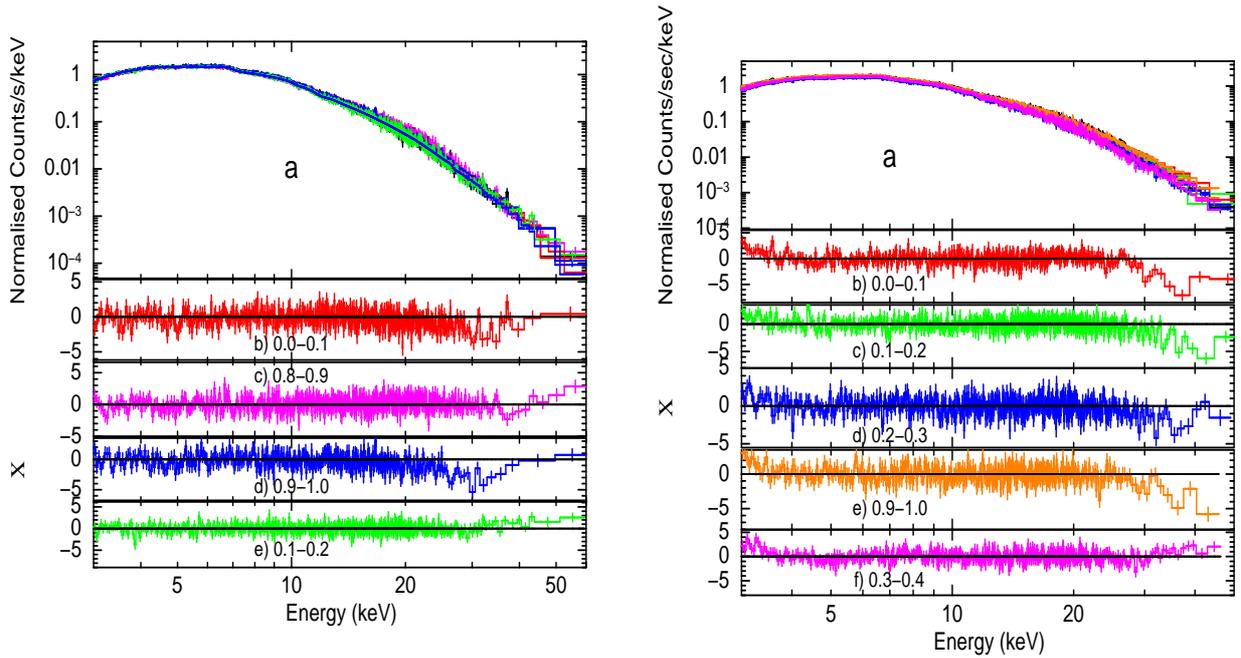

\begin{minipage}{0.45\textwidth}
\includegraphics[height=\columnwidth,width=\columnwidth,angle=-90]{fig17.eps}
\end{minipage}
\hspace{0.02 \linewidth}
\begin{minipage}{0.45\textwidth}
\includegraphics[height=\columnwidth,width=1.1\columnwidth,angle=-90]{fig18.eps}
\end{minipage}
        \caption{a.~We show phase-resolved spectra of 4U~1901+03 during Obs~1(left) \& 2(right).~For clarity, we have included data only from \textsc{FPMA}.~Panels (b,c,d) in the left show residuals from the best-fit phase-resolved spectra of 4U 1901+03, obtained using Model~2 without additional absorption component at $30~\rm{keV}$ and for comparison we show residuals of the phase-resolved spectrum where no negative residuals around $30~\rm{keV}$ were observed (bottom panel,~e).~Similarly in the right, panels~(b,c,d,e)~indicate negative residuals around $E_{cyc}$ while the bottom panel~(f) are residuals across the pulse phase where $30~\rm{keV}$ feature was not seen.}
\label{Ecyc-spec-1-2}
\end{figure*}

The 30~keV absorption feature was not detected in the phase-averaged spectra of Obs~1 and Obs~2.
In order to probe the presence of this feature, we looked carefully at the pulse-phase resolved spectra of these observations.~Consequently, for Obs~1, we found dip-like structures at around 30~$\rm{keV}$, 38~$\rm{keV}$, and 29~$\rm{keV}$ in the spectral residuals of the following three pulse phase~0.0--0.1,~0.8--0.9,~0.9--1.0,~respectively.~This feature did not appear in the rest of the pulse phases.~Similarly,~for Obs~2, we found the presence of $E_{cyc}$ at around 38~$\rm{keV}$ in the phase-resolved spectra at 0.0--0.1,~0.1--0.2 pulse phase,~34~$\rm{keV}$ for the spectrum at 0.2--0.3 pulse phase, and 40~$\rm{keV}$ for 0.9-1.0 pulse phase~(Figure~\ref{Ecyc-spec-1-2}).~These values of $E_{cyc}$ are consistent with those observed for Obs~3 and Obs~4.~Therefore, an additional `\textit{GABS}' component was added to the X-ray spectra of these observations.~We computed the statistical significance of detection of the CRSF using the run-test as opted for phase-averaged spectra of Obs~3 and Obs~4~(refer to Section~4.1 for details).~The values mentioned in Table~\ref{tab:significance} and the residuals in Figure~\ref{Ecyc-spec-1-2} show that the detection of the CRSF in the X-ray spectra at certain pulse phases is significant.~The variation of $E_{cyc}$ parameters with the pulse phase is shown in Figure~\ref{Phase-resol-1-2}.~During Obs~1, $E_{cyc}$ was observed during the off-peak pulse phases, whereas  Obs~2, the CRSF was detected during the pulse peaks.~${\tau}_{cyc}$ showed a trend similar to that observed for $E_{cyc}$ during both observations.
~The values of X-ray flux across pulse phases where the CRSF has been detected were compared to those measured during phases where there is no detection.~Thus,~presence/absence of CRSF did not show a clear dependence on the X-ray flux.

\begin{table*}
\centering

	\caption{Table shows the values of probability of chance improvement~(PCI) obtained on performing the run-test on residuals of the phase-resolved X-ray spectra of Obs~1 and Obs~2.~PCI corresponds to $N_r<=3$~(number of runs), $N_+$~(points above zero) and $N_-$~(points below zero). }
        \label{tab:significance}
	\begin{tabular}{|c|c|l|c|c|}
        \hline

		  &    Obs~1  & &    &    \\
        \hline
		Pulse Phase     &  $PCI$  & $N_+$    & $N_-$   \\
        \hline \hline

		0.0-0.1        & 0.3~$\%$ & 6    &  8              \\
		0.8-0.9        & 0.05$\%$ & 7    &  11  \\
		0.9-1.0        & 0.02$\%$ & 16    &  5 \\
    
\hline
		&      Obs~2   & &   &   \\
        \hline
		Pulse Phase     & $PCI$  & $N_+$   & $N_-$    \\
        \hline \hline

		0.0-0.1 &       0.09$\%$     &  9   &  7         \\
		0.1-0.2 &       0.03$\%$     &  9   &  9          \\
		0.2-0.3 &       0.04$\%$     &  3   &  22          \\
		0.9-1.0 &       0.5$\%$      &  2   &  21           \\
\hline \hline

	\end{tabular}
         \flushleft
 
\end{table*}

\section{Discussion}
In this work, we have performed a detailed broadband timing and spectral analysis with \emph{Swift} and \emph{NuSTAR}.~We discuss our results as follows:
\subsection{Timing Results}
\subsubsection{Pulse Profile Evolution with the X-ray Luminosity}
As can be seen from Figure~\ref{outburst} that \emph{NuSTAR} and \emph{Swift} observations were taken at different levels of intensity and, therefore, we calculated the values of X-ray luminosity~($L_X$) during these observations~(Obs~1,~2,~3 \& 4),~assuming source distance to be $3~{\rm kpc}$.~We have used the 
values of X-ray flux obtained in the 1--70~$\rm{keV}$ band during these observations~(refer to Tables~\ref{spec-para12} \& \ref{spec-para34}) and found $L_X$
to be $\sim$ $0.76{\times}10^{37} {\rm ergs~s^{-1}}$, $0.85{\times}10^{37} {\rm ergs~s^{-1}}$, $0.5{\times}10^{37} {\rm ergs~s^{-1}}$, $0.3{\times}10^{37} {\rm ergs~s^{-1}}$ for Obs~1,~2,~3,~\& 4, respectively.~Thus, these observations allowed us to probe into luminosity dependence of pulse profiles. We found that pulse profiles show a strong evolution with the change in $L_X$ (see Figure~\ref{Avg-Obs12}).\\

Due to effect of the non-spherical emission region
and scattering cross sections of the photons which are altered in the presence of strong magnetic fields,~different beaming patterns of radiation, ``Fan" or ``Pencil"
are produced depending on the mass accretion rate on the neutron star \citep{Nagel81, Meszaros85}.~The critical luminosity~($L_{crit}$) indicates whether the radiation pressure of the emitting plasma is capable of decelerating the accretion flow and plays a role in defining two accretion regimes.~If the $L_X$ is greater than $L_{crit}$ (super-critical regime), then radiation pressure is high enough to stop the accreting matter at a distance above the neutron star, forming a radiation-dominated shock~(Fan beam). If $L_X$ is less than $L_{crit}$, then the accreted material reaches the neutron star surface through coulomb collisions with thermal electrons or through nuclear collisions with atmospheric protons~(Pencil beam)~\citep{Harding94}.~However, it has been found/proposed that beaming patterns can be much more complex than a simple pencil/fan beam \citep[][]{Kraus95,Kraus96,Becker12,Mushtukov2015} and accretion regimes can also be probed by observing changes in the pulse profiles, cyclotron line energies, changes in the spectral shape \citep[see e.g.,][]{Parmar89,Reig13,Mushtukov2015}.~Observations made with \emph{NICER} and \emph{Insight-HXMT} revealed the value of critical luminosity in this source to be $\sim$~$1{\times}10^{37} {\rm ergs~s^{-1}}$~\citep{Ji2020,Tuo2020}.~Thus,~observations used in our work were taken below the critical luminosity or belong to sub-critical regime. \\

From Figure~\ref{Avg-Obs12}, we found a strong luminosity dependence of the pulse profiles of 1901.~A secondary peak~(or notch) could be clearly seen in the pulse profiles of Obs~1 \& 2~(except \emph{Swift} pulse profiles) while the pulse profiles during Obs~3 \& 4 were simple, broad and single-peaked.~\citet{Ji2020} found that close to $0.8{\times}10^{37} {\rm ergs~s^{-1}}$, the notch (or secondary peak) appears in the pulse profiles, whereas below $0.7{\times}10^{37} {\rm ergs~s^{-1}}$ pulse profiles are single peaked.~Thus, our results are consistent with that obtained by \citet{Ji2020} and similar pulse profiles were also seen during the 2003 outburst of 1901 \citep[see e.g.,][]{Chen08,Reig16}.~Based on a detailed study of pulse profiles of 1901 using the \emph{NICER} and \emph{Insight-HXMT} data,~\citet{Ji2020} proposed that pulse profiles of 1901 show a complex luminosity dependence
and can be formed with a combination of ``fan" and ``pencil" beams.~Fan beam dominating the high luminosity observations while pencil beam at low luminosity \citep[see also][]{Chen08}. \\

\subsubsection{Energy dependence of pulse profiles: An abrupt change close to the CRSF}

A strong energy dependence of pulse profiles has been observed in several X-ray pulsars like 4U~0115+63~\citep{Tsygankov07}, 4U~1626--67~\citep[see e.g.,][]{Beri14}, LMC~X--4 \citep[see e.g.,][]{Beri17},~1A 1118-61 \citep[e.g.,][]{Maitra12}.~It is usually found that pulse profiles tend to be simpler at higher energies~(above $10~\rm{keV}$) compared to complex profiles observed at lower energies \citep[see e.g.,][]{Devasia11}.~A similar energy dependence of the pulse profiles was found during Obs~1 and 2 (see Figure~\ref{EN-pp1} \& \ref{EN-pp2}).~However, it was interesting to note that during Obs~3, energy-resolved pulse profiles showed a change in the general trend. They changed from simple, broad to complex pulse profiles above 28 keV.~Two prominent peaks with a dip at around pulse phase 0.5 appeared in the pulse profiles of 28-37~$\rm keV$ energy band (see Figure~\ref{EN-pp3}).~Obs~4 also showed similar features in the pulse profile of this energy band.~Based on the relationship proposed by \citet{Mushtukov15} between the critical luminosity and magnetic field of neutron star,~\citet{Ji2020} suggested 30~$\rm keV$ to be energy of the CRSF in 1901. 
Scattering cross sections are believed to be significantly modified and increase by a large factor near the CRSF \citep{Araya99, Araya2000, Schonherr07}, causing a change in the beaming patterns.
There also exist several reports where significant changes in the shape of pulse profiles have been found near the cyclotron line energies see e.g., V~0332+53 \citep{Tsygankov06}, 4U~0115+63~\citep{Ferrigno11}.
Thus, it may be possible that the abrupt change observed in the pulse profiles around 30~$\rm keV$ is due to the existence of cyclotron line of this system at around this energy. 
Such prominent changes in the pulse profiles were only seen in observations made during declining phase of the outburst~(obs~3 \& 4), this further suggest a strong luminosity dependence of the beaming pattern. \\

Pulse fraction which quantifies the fraction
of X-ray photons contributing to the observed pulsation showed an increase with the increasing energy, suggesting large contribution from the hard X-rays.~Similar dependence on energy has also been observed in several X-ray pulsars~\citep{Nagase89,Bildsten1997}.  \\

\subsection{Spectral Results}

\subsubsection{$L_X$ Dependence: Continuum parameters,~10~$\rm{keV}$ feature and iron line }
Our spectral results indicated the presence of hard X-ray spectrum which is typical of X-ray pulsars.~The values of $\Gamma$ found are consistent with that observed during its previous
outburst in 2003 \citep[see Figure~4 of][]{Reig16}.~The ``10 $\rm {keV}$ feature" observed in the X-ray spectra during the 2019 outburst showed a trend which is similar to that observed during the 2003 outburst of 1901.~We discuss the origin of this feature later in this section.~A positive correlation was observed in the energy of this feature with the X-ray flux as observed earlier \citep[see Figure~7 of][]{Reig16}.~Iron line energies showed similar values during all four observations, however, equivalent width showed a higher value during Obs~2, indicating luminosity dependence.~This is again consistent with previous reports on this source \citep{Reig16}.\\

\subsubsection{Cyclotron line at $30~{\rm{keV}}$:~Magnetic Field Estimate, Pulse-phase dependence and transient nature}

One of the significant outcome of this study is the confirmation of cyclotron line~($E_{cyc}$) around 30 $\rm {keV}$ in the X-ray spectra of 1901 and studying its variable nature both with luminosity and pulse phase.~This feature was first detected by \citet{Coley19}, however, this is for the first time the existence of this feature has been well established.~An abrupt change in the pulse profiles around this energy was observed.~There have also been other sources where an abrupt change in the pulse shape near the CRSF energy has been observed, as an example V0332+53 \citep[see][for details]{Tsygankov06}.
The presence of a broad absorption feature in the X-ray continuum of X-ray pulsars is reminiscent of a cyclotron line absorption feature and its energy is related to the magnetic field of neutron star:
$E_{cyc}$~=~$11.6/(1+z)~B_{12}~{\rm keV}$ where $B_{12}$ is the magnetic field strength in units of $10^{12}~\rm{G}$ and z is gravitational redshift~(0.3 is a typical value used for standard neutron star parameters).~Using this relation, we inferred magnetic field of the neutron star to be $\sim$~$3.5\times10^{12} \rm{G}$.\\


CRSF at $30~{\rm{keV}}$ in 1901 was detected in the phase-averaged spectra of only low luminosity observations, and and at certain phases during the bright observations indicating it to be a transient feature.~\citet{Coley19} also indicated the presence of an additional narrow absorption feature around 30~$\rm{keV}$ in the \emph{NuSTAR} spectra during the declining phase of the outburst which was not seen in the observations made during the rising phase of the outburst. \\

Cyclotron line observed in the X-ray spectra originate from the accretion column \citep{Nishimura14,Schonherr14}, which is highly luminosity dependent. It has also been proposed in some cases that the line can be formed as a 
result of the reflection of X-rays from the atmosphere of the neutron star \citep{Poutanen13,Lutovinov15}.
Non-detection of CRSF at certain pulse phases is explained as a result of large gradient of the magnetic field strength over the visible column height or latitudes on the stellar surface \citep[see, e.g.][]{Molkov19}.
Thus,~there may be possibility that during Obs~1 and Obs~2, the $30~{\rm{keV}}$ feature could be smeared out in the phase-averaged spectrum due to its large variability across the phase, and the possible effects due to light bending \citep{Falkner2018}. \\

An important characteristic of CRSF is its dependence on the pulse phase.
Our results showed a strong dependence of 30~$\rm{keV}$ feature on the pulse-phase.
$E_{cyc}$ varied by $\sim$~$60\%$ and $\sim$~$47\%$ during Obs~3 and Obs~4, respectively.
The CRSF detected at certain phases of Obs~1 and Obs~2 also showed values similar to that 
observed during Obs~3 and Obs~4.~A similar variation of CRSF has also been observed in other sources.~Such variation of CRSF is understood to be a result of sampling different heights of the line forming region as a function of pulse phase or alternatively a complex accretion geometry or a large gradient in the B field indicating a non-dipolar geometry~\citep[see][for details]{Maitra17b, Staubert2019}.~From Figure~\ref{tau-pp} it is also evident that optical depth shows a trend similar to the pulse profile in the energy band coinciding with the band where CRSF is identified in the spectrum of this source.~Thus,~this must be taken into account while constructing models of CRSFs in the neutron star atmosphere.

\subsubsection{Pulse-phase dependence of spectral parameters}
From our pulse phase-resolved spectroscopy, we observed a negative correlation of photon
index with the X-ray flux.~Our hardness ratio plots also showed a correlation between intensity and hardness (Figure~\ref{HR1}).~Thus,~our spectral results are in consistent with our model-independent approach to probe into the spectral evolution across the pulse phase.~We observed an anti-correlation between the iron line EW and pulse profiles
which may suggest that the amount of neutral matter obscuring the neutron star is higher at the off-pulse phases that gives rise to stronger iron fluorescence lines.~We observed a strong pulse phase dependence of ``10 $\rm {keV}$ feature" which is consistent with that observed during previous outburst of 1901 in 2003
\citep{Chen08,Reig16}.
However, it is also believed that ``10 $\rm {keV}$ feature" can also arise due to the limitation of simple phenomenological models used for describing the X-ray continuum \citep[see][]{Coburn02, Staubert2019}.~Moreover,~our timing results~(pulse profiles) did not show any 
significant deviation from the general trend around this energy~($10~\rm{keV}$).
This further support our speculation that the observed ``10 $\rm {keV}$ feature"
may have a non-magnetic origin. 

\begin{table*}
\caption{Spectral fit parameters with other phenomenological models.~We have performed spectral fitting  in the 1-79~keV band. }
\label{spec-para12}
\begin{tabular}{ c c c c c c c c c}
\hline
\hline
Parameters & \text{Model 1} & \text{Model 2} &  \text{Model 3} & \text{Model 4}  & \text{Model 5} & \text{Model 6}  \\
\hline
&   &  Observation~1 &  \\
\hline
	$N\rm{_{H}}$ ($10^{22}\rm{cm^{-2}}$) &  $4.5\pm0.2$ & $5.6\pm{0.2}$ &  $6.2\pm0.3$  &  $5.3\pm0.2$           & $7.2\pm0.3$            & $8.9\pm0.2$ \\ [0.1cm]
 
$\Gamma$                              & $0.55\pm0.02$  & $0.60\pm{0.02}$           &  $0.53\pm0.03$  &  $0.58\pm0.02$                             &        $1.3\pm0.1$             & -                        \\ [0.1cm]

$E_{cut/break/e}$ (keV)              &  $3.65\pm0.07 $ &  $7.6\pm{0.1}$           &  $4.73\pm0.04$  & $1.7\pm0.2$            &      $10\pm2$                          & -                           \\ [0.1cm]

$E_{fold}$ (keV)                     &  $7.5\pm0.1 $  & -                          &-                            &  $7.5\pm0.1$           &       $8.4\pm0.1$                           & -                            \\ [0.1cm]

$kT\rm_{bb}~(keV)$                   & $0.17\pm0.02$    & $0.15\pm0.01$             & $0.15\pm0.01$             & $0.15\pm0.01$           & $0.14\pm0.01$                      & $0.131\pm0.005$    \\ [0.1cm]

$kT\rm_{e}~(keV)$                   & -   & -             & -             & -           & -                     & $4.94\pm0.03$    \\ [0.1cm]

$\tau$                               & -  &  - & - & - & -  & $4.92\pm0.04$             \\ [0.1cm]
	$N^{a}$                              & $0.21\pm0.01$ & $0.39\pm{0.01}$ & $0.46\pm0.02$  &     $0.29\pm0.02$     &      $0.92\pm0.04$ & -     \\ [0.1cm]

$E_{Fe}$ (keV) & $6.52\pm0.03 $ & $6.52\pm0.03$ & $6.50\pm0.03$ & $6.52\pm0.03$ & $6.51\pm0.03$ & $6.6\pm0.1$ \\[0.1cm]

$EW_{Fe}$ (eV)  & $57\pm7$ &  $48\pm6$ & $50\pm7$ &  $52\pm8$ & $49\pm7$ & $20\pm3$ \\ [0.1cm]

 $E_{Gabs\_add}$ (keV)         &  $10.8\pm0.1$   &  $10.7\pm{0.1}$ &   $10.8\pm0.2$          &    $10.7\pm0.1$   &  $11.3\pm0.1$         & $11.3\pm0.3$   \\ [0.1cm]

$\sigma_{Gabs\_add}$ (keV)  &    $3.4\pm0.3$  &  $3.6\pm0.3 $ & $1.5\pm0.2$                             & $3.8\pm0.3$           &              $0.9\pm0.2$                                & $1.69\pm0.04$   \\ [0.1cm]
    
${\tau}_{Gabs\_add}$     &    $0.16\pm0.05$  & $0.17\pm{0.05}$   & $0.15\pm0.04$                           & $0.17\pm0.05$           &              $0.22\pm0.03$             &   $0.08\pm0.03$  \\ [0.1cm]

$const_{FPMB}$                      & $0.978\pm0.002$               & $0.978\pm0.002$               & $0.978\pm0.002$             &  $0.978\pm0.002$                    & $0.978\pm0.002$                     & $0.978\pm0.002$               \\ [0.1cm]

$const_{XRT}$                       & $0.99\pm0.01$               & $0.99\pm0.01$               &  $0.99\pm0.01$             & $0.99\pm0.01$                    & $1.00\pm0.01$                     & $0.99\pm0.01$                \\ [0.1cm]
${\text{Unabsorbed Flux (1-79 keV)}}^b$  & $7.1\pm0.1$ & $7.1\pm0.1$ & $7.1\pm0.1$ & $7.1\pm0.1$ & $7.1\pm0.1$ & $7.1\pm0.1$ \\ [0.1cm]
Reduced ${\chi}^2$~(dof)            & 1.24~(2223)                   &  1.28~(2224)                & 1.27~(2221)                  & 1.27~(2221)                      &        1.27~(2221)                        & 1.31~(2222)                   \\ [0.1cm]
\hline
&   & Observation~2 &  \\
\hline
 $N\rm{_{H}}$ ($10^{22}\rm{cm^{-2}}$) &  $4.3\pm0.3$ & $5.7\pm0.2$ &  $5.2\pm0.2$  &  $5.3\pm0.3$     & $6.4\pm0.2$    & $1.82\pm0.07$ \\ [0.1cm]
 
$\Gamma$                              & $0.51\pm{0.03}$  & $0.56\pm0.02$     &  $0.41\pm0.06$  &  $0.55\pm0.03 $       &    $0.93\pm0.03$             & -                         \\ [0.1cm]

$E_{cut/break/e}$ (keV)              &  $3.8\pm0.1 $ &  $7.1\pm0.1$           &  $6.2\pm0.3$  & $1.9\pm0.5$            &      $1^c$                          & -                           \\ [0.1cm]

$E_{fold}$ (keV)                     &  $6.9\pm0.1 $  & -                          &-                            &  $6.97_{-0.07}^{+0.09}$           &       $7.5\pm0.1$                           & -                            \\ [0.1cm]

$kT\rm_{bb}~(keV)$                   & $0.18\pm0.01$    & $0.15\pm0.01$             & $0.15\pm0.01$      & $0.16\pm0.01$           & $0.15\pm0.01$                      & -   \\ [0.1cm]

$kT\rm_{e}~(keV)$                   & -   & -             & -             & -           & -                     & $4.69\pm0.04$    \\ [0.1cm]

$\tau$                               & -  &  - & - &   - & - & $4.85\pm0.04$             \\ [0.1cm]
$N^{a}$                              &  $0.23\pm0.01$ & $0.44\pm0.01$ & $0.37\pm0.02$ &   $0.32\pm0.02$ & $1.07\pm0.04$ &      -             \\ [0.1cm]

$E_{Fe}$ (keV) & $6.50\pm0.03 $ & $6.53\pm0.03$ & $6.52\pm0.03$ & $6.53\pm0.03$ & $6.55\pm0.04$ & $6.53\pm0.03$ \\[0.1cm]

$EW_{Fe}$ (eV) & $97\pm10$ & $75\pm8$& $66\pm7$ & $80\pm9 $& $98\pm9$ & $75\pm9$ \\ [0.1cm]

$E_{Gabs\_add}$ (keV)         &  $11.4\pm0.1$   &  $11.4\pm0.1$ & $ 11.3\pm0.1$     & $11.4\pm0.1$      &   $11.6\pm0.1$    & $10.7\pm0.3$   \\ [0.1cm]

$\sigma_{Gabs\_add}$ (keV)  &   $1.9\pm0.2$ &  $1.9\pm0.2$ & $2.2\pm0.2$           & $1.9\pm0.2$         &   $1.8\pm0.2$    & $1.4\pm0.2$   \\ [0.1cm]
    
${\tau}_{Gabs\_add}$     &    $0.09\pm0.01$  & $0.09\pm0.01$   & $0.11\pm0.02$  & $0.09\pm0.01$  &    $0.09\pm0.01$    &   $0.04\pm0.01$  \\ [0.1cm]

$const_{FPMB}$                      & $1.080\pm0.003$               & $1.081\pm0.003$               & $1.080\pm0.003$             &  $1.080\pm0.003$                    & $1.080\pm0.003$                     & $1.080\pm0.003$               \\ [0.1cm]

$const_{XRT}$                       & $1.048_{-0.008}^{+0.009}$               & $1.04\pm0.01$               &  $1.040_{-0.008}^{+0.009}$             & $1.04\pm0.01$                   & $1.041\pm0.007$               & $1.044\pm0.008$              \\ [0.1cm]

${\text{Unabsorbed Flux (1-79 keV)}}^b$ & $8.0\pm0.2$ & $8.1\pm0.1$ & $8.1\pm0.2$ & $8.1\pm0.2$ & $8.0\pm0.2$ & $8.0\pm0.2$ \\ [0.1cm]

Reduced ${\chi}^2$~(dof)            & 1.12~(2140)        &  1.14~(2141)        & 1.13~(2140)    & 1.14~(2140)    &        1.17~(2141)                        & 1.18~(2142)                   \\ [0.1cm]
\hline
\end{tabular}
\\
\begin{flushleft}
{{\bf{Note}}:  
{a $\rightarrow$ Normalization~($N_{}$)
     is in units of $\rm{photons~cm}^{-2}~\rm{s}^{-1}~\rm{keV}^{-1}$ at 1~keV.} \\
{b $\rightarrow$ Unabsorbed flux in units $10^{-9} ergs \, cm ^{-2} s^{-1}$
 } \\
{c $\rightarrow$ fixed parameters}
}

Model 1: const*tbabs*(powerlaw*highecut*gabs+bbodyrad+gauss) \\
Model 2: const*tbabs*(cutoffpl*gabs+bbodyrad+gauss) \\
Model 3: const*tbabs*(NPEX*gabs+bbodyrad+gauss) \\
Model 4: const*tbabs*(powerlaw*newhcut*gabs+bbodyrad+gauss)\\
Model 5: const*tbabs*(powerlaw*fdcut*gabs+bbodyrad+gauss)\\
Model 6: const*tbabs*(comptt*gabs+bbodyrad+gauss) \\
\end{flushleft}
\end{table*}

\begin{table*}
\caption{Spectral fit parameters obtained for Observation~3 and 4.}
\label{spec-para34}
\begin{tabular}{ c c c c c c c c c}
\hline
\hline
Parameters & \text{Model 1} & \text{Model 2} &  \text{Model 3} & \text{Model 4}  & \text{Model 5} & \text{Model 6}  \\
\hline
&   &  Observation~3 &  \\
\hline
 $N\rm{_{H}}$ ($10^{22}\rm{cm^{-2}}$) &  $3.4\pm0.1$ & $4.4\pm0.2$ &   $3.8\pm0.1$  &  $3.9\pm0.2$  & $5.6\pm0.3$  & $6.7\pm0.2$ \\ [0.1cm]
 
$\Gamma$     & $0.55\pm0.02$  & $0.61\pm0.04$    &  $0.46\pm0.05$  &  $0.56\pm0.04$  &  $1.3\pm0.1$  & -                        \\ [0.1cm]

$E_{cut/break/e}$ (keV)  &  $3.2\pm0.1 $ &  $7.4\pm0.2$   &  $4.8\pm0.1$  & $1.1\pm0.3$  &    $11\pm2$    & -                           \\ [0.1cm]

$E_{fold}$ (keV)    &  $7.2\pm0.1 $  & -   &-   &  $7.44_{-0.22}^{+0.16}$  &   $8.0\pm0.2$  & -                            \\ [0.1cm]

$kT\rm_{bb}~(keV)$  & $0.08\pm0.02$    & $0.10\pm0.02$  & $0.12_{-0.02}^{+0.01}$   & $0.10\pm0.02$   & $0.12\pm0.01$     & $0.12\pm0.01$    \\ [0.1cm]

$kT\rm_{e}~(keV)$    & -   & -             & -             & -           & -                     & $5.0\pm0.1$    \\ [0.1cm]

$\tau$        & -  &  - & - & - & -  & $5.1\pm0.1$             \\ [0.1cm]
$N^{a}$     & $0.158\pm0.005$ & $0.28\pm0.01$ & $0.27\pm0.02$  &   $0.23\pm0.01$     &  $0.63\pm0.03$ & -     \\ [0.1cm]

$E_{Fe}$ (keV) & $6.54\pm{0.03} $ & $6.54\pm0.03$ & $6.53\pm0.03$ & $6.55\pm0.03$ & $6.52\pm0.04$ & $6.45\pm0.05$ \\[0.1cm]

$EW_{Fe}$ (eV)  & $64\pm9$ &  $64\pm9$ & $58\pm8$ &  $64\pm8$ & $61\pm8$ & $116\pm16$ \\ [0.1cm]

$E_{Gabs\_add}$ (keV)  &  $10.2\pm0.2$   &  $9.9\pm0.2$ &   $10.4\pm0.1$   &    $10.0\pm0.2$   &  $10.9\pm0.1$ & $11.5\pm0.2$   \\ [0.1cm]

$\sigma_{Gabs\_add}$ (keV)  &   $3.2\pm0.3$  &  $3.7\pm0.4$ & $2.3\pm0.3$  & $3.6\pm0.4$  &   $1.9\pm0.3$   & $1.7\pm0.3$   \\ [0.1cm]
    
${\tau}_{Gabs\_add}$     &    $0.17\pm0.02$  & $0.18\pm0.04$   & $0.08\pm0.01$  & $0.18\pm0.02$  &  $0.07\pm0.02$  &   $0.07\pm0.01$  \\ [0.1cm]

$E_{cyc}$ (keV)         &  $32.4\pm0.7$   & $33\pm2$ &   $33\pm1$  &    $33\pm2$   &  $32\pm1$  & $34\pm1$   \\ [0.1cm]

$\sigma_{cyc}$ (keV)  &    $4.8\pm0.5$  & $6\pm1$ & $7.0\pm2.0$ & $5\pm1$ &   $5\pm1$ &  $9.96^c$\\ [0.1cm]
    
${\tau}_{cyc}$     &    $0.21\pm0.06$  & $0.23\pm0.08$  & $0.4\pm0.1$    & $0.23\pm0.09$    &  $0.23\pm0.09$  &   $0.48\pm0.07$  \\ [0.1cm]

$const_{FPMB}$                      & $1.052\pm0.003$               & $1.052\pm0.003$            & $1.052\pm0.003$           &  $1.052\pm0.003$                    &$1.052\pm0.003$                     & $1.052\pm0.003$               \\ [0.1cm]

$const_{XRT}$                       & $1.15\pm0.01$               & $1.15\pm0.01$              &  $1.15\pm0.01$            & $1.15\pm0.01$                   & $1.15\pm0.01$                    & $1.16\pm0.01$               \\ [0.1cm]
${\text{Unabsorbed Flux (1-79 keV)}^b}$ & $4.7\pm0.1$ & $4.7\pm0.1$ & $4.7\pm0.1$ & $4.7\pm0.1$ & $4.7\pm0.1$ & $4.7\pm0.1$ \\ [0.1cm]
Reduced ${\chi}^2$~(dof)    & 1.10~(2064)   &  1.14~(2064)  & 1.16~(2064)     & 1.14~(2063)   & 1.12~(2063) &        1.14~(2065)       \\ [0.1cm]
\hline
&   & Observation~4 &  \\
\hline
 $N\rm{_{H}}$ ($10^{22}\rm{cm^{-2}}$) &  $3.1\pm0.1$ & $3.7\pm0.1$ &  $4.8\pm0.2$  &  $3.3\pm0.2$  & $5.1\pm0.2$            & $5.5\pm0.2$ \\ [0.1cm]
 
$\Gamma$    & $0.62\pm0.03$  & $0.68\pm0.02$  &  $0.87\pm0.02$  &  $0.63\pm0.02$  &  $1.44\pm0.05$ & -                         \\ [0.1cm]

$E_{cut/break/e}$ (keV)   &  $3.1\pm0.1 $ &  $7.9\pm0.1$  &  $4.63\pm0.05$  & $1.2\pm0.3$  &  $17\pm{2}$                          & -                           \\ [0.1cm]

$E_{fold}$ (keV) &  $7.6\pm0.1 $  & -  &-  &  $7.6\pm0.1$ & $7.7\pm0.1$     & -    \\ [0.1cm]

$kT\rm_{bb}~(keV)$   & $0.08\pm0.02$   & $0.08\pm0.02$  & $0.10\pm0.01$ & $0.09\pm0.01$ & $0.10\pm0.01$                      &  $0.10\pm0.01$  \\ [0.1cm]

$kT\rm_{e}~(keV)$    & -   & -             & -             & -           & -  & $4.94\pm0.06$    \\ [0.1cm]

$\tau$              & -  &  - & - &   - & - & $5.18\pm0.06$             \\ [0.1cm]
$N^{a}$           &  $0.13\pm0.02$ & $0.19\pm{0.01}$ & $0.30\pm0.01$ & $0.14\pm0.02$ & $0.38\pm0.02$ &      -             \\ [0.1cm]

$E_{Fe}$ (keV) & $6.57\pm0.03 $ & $6.58\pm0.03$ & $6.54\pm0.03$ & $6.54\pm{0.03}$ & $6.56\pm0.03$ & $6.55\pm0.03$ \\[0.1cm]

$EW_{Fe}$ (eV) & $43\pm6$ & $42\pm6$ & $51\pm6$ & $41_{-30}^{+17}$ & $46\pm7$ & $44\pm6$\\ [0.1cm]

$E_{Gabs\_add}$ (keV) &  $8.8\pm0.1$  &  $8.6\pm0.1$ & $ 11.2\pm0.6$  & $9.1\pm0.3$  &   $9.1\pm0.3$   & $10.2\pm0.3$   \\ [0.1cm]

$\sigma_{Gabs\_add}$ (keV)  &   $4^c$ &  $4^c$ & $1.3\pm0.6$  & $4^c$ & $5\pm1$ & $3.6\pm0.5$  \\ [0.1cm]
    
${\tau}_{Gabs\_add}$     &    $0.29\pm0.02$  & $0.29\pm0.02$   & $0.02\pm0.02$ & $0.28\pm0.02$  &  $0.13\pm0.02$                                 &   $0.09\pm0.03$  \\ [0.1cm]

$E_{cyc}$ (keV)         &  $32.6\pm0.8$   &  $35\pm1$ & $ 31.7\pm0.6$   & $32\pm1$  &  $30\pm1$ & $32\pm1$   \\ [0.1cm]

$\sigma_{cyc}$ (keV)  &    $3.45^c$  & $4.44^c$ & $6.73^c$ & $3.5^c$ &  $4.12^c$  &  $5.54^c$\\ [0.1cm]

${\tau}_{cyc}$     &    $0.14\pm0.03$  & $0.18\pm0.03$   & $0.31\pm0.04$  & $0.12\pm0.03$ &  $0.16\pm0.03$                                 &   $0.23\pm0.02$  \\ [0.1cm]

$const_{FPMB}$                      & $1.025\pm0.002$               & $1.025\pm0.002$               & $1.025\pm0.002$             &  $1.025\pm0.002$                    & $1.025\pm0.002$                     & $1.025\pm0.002$               \\ [0.1cm]

$const_{XRT}$                       & $0.94\pm0.01$               & $0.94\pm0.01$               &  $0.94\pm0.01$            & $0.94\pm0.01$                   & $0.94\pm0.01$              & $0.94\pm0.01$            \\ [0.1cm]

${\text{Unabsorbed Flux (1-79 keV)}^b}$  & $2.9\pm0.1$ & $2.9\pm0.1$ & $2.9\pm0.1$ & $2.9\pm0.1$ & $2.9\pm0.1$ & $2.9\pm0.1$ \\ [0.1cm]

Reduced ${\chi}^2$~(dof)     & 1.13~(2064)    &  1.17~(2065)  & 1.09~(2063) & 1.17~(2064)  &        1.09~(2064)                        & 1.08~(2064)                   \\ [0.1cm]
\hline
\end{tabular}
\\

{{\bf{Note}}:  
{a $\rightarrow$ Normalization~($N_{}$)
     is in units of $\rm{photons~cm}^{-2}~\rm{s}^{-1}~\rm{keV}^{-1}$ at 1~keV.} \\
{b $\rightarrow$ Unabsorbed flux in units $10^{-9} ergs \, cm ^{-2} s^{-1}$
 } \\
 {c $\rightarrow$ fixed parameters} \\
}

\end{table*}

\section*{Acknowledgments}
The authors gratefully acknowledge the referee for his/her useful suggestions that helped us to improve the presentation of the paper.
 A.B is grateful to the Royal Society and SERB (Science and Engineering Research Board, India).~A.B. is supported by an INSPIRE Faculty grant (DST/INSPIRE/04/2018/001265) by the Department of Science and Technology, Govt. of India.~She would also like to thank Prof. Biswajit Paul for fruitful discussions.
This research has made use of data and/or software
provided by the High Energy Astrophysics Science Archive Research
Center (HEASARC), which is a service of the Astrophysics Science
Division at NASA/GSFC and the High Energy Astrophysics Division of the
Smithsonian Astrophysical Observatory.~This research has made use of
NASA's Astrophysics Data System.

\section{DATA AVAILABILITY}
The data underlying this article
are publicly available from  the  High  Energy  Astrophysics  Science Archive Research Center (HEASARC), 
provided by NASA's Goddard Space Flight Center.

\bibliographystyle{mnras}
\bibliography{ref}

\begin{thebibliography}{}
\makeatletter
\relax
\def\mn@urlcharsother{\let\do\@makeother \do\$\do\&\do\#\do\^\do\_\do\%\do\~}
\def\mn@doi{\begingroup\mn@urlcharsother \@ifnextchar [ {\mn@doi@}
  {\mn@doi@[]}}
\def\mn@doi@[#1]#2{\def\@tempa{#1}\ifx\@tempa\@empty \href
  {http://dx.doi.org/#2} {doi:#2}\else \href {http://dx.doi.org/#2} {#1}\fi
  \endgroup}
\def\mn@eprint#1#2{\mn@eprint@#1:#2::\@nil}
\def\mn@eprint@arXiv#1{\href {http://arxiv.org/abs/#1} {{\tt arXiv:#1}}}
\def\mn@eprint@dblp#1{\href {http://dblp.uni-trier.de/rec/bibtex/#1.xml}
  {dblp:#1}}
\def\mn@eprint@#1:#2:#3:#4\@nil{\def\@tempa {#1}\def\@tempb {#2}\def\@tempc
  {#3}\ifx \@tempc \@empty \let \@tempc \@tempb \let \@tempb \@tempa \fi \ifx
  \@tempb \@empty \def\@tempb {arXiv}\fi \@ifundefined
  {mn@eprint@\@tempb}{\@tempb:\@tempc}{\expandafter \expandafter \csname
  mn@eprint@\@tempb\endcsname \expandafter{\@tempc}}}

\bibitem[\protect\citeauthoryear{{Araya} \& {Harding}}{{Araya} \&
  {Harding}}{1999}]{Araya99}
{Araya} R.~A.,  {Harding} A.~K.,  1999, \mn@doi [\apj] {10.1086/307157}, \href
  {https://ui.adsabs.harvard.edu/abs/1999ApJ...517..334A} {517, 334}

\bibitem[\protect\citeauthoryear{{Araya-G{\'o}chez} \&
  {Harding}}{{Araya-G{\'o}chez} \& {Harding}}{2000}]{Araya2000}
{Araya-G{\'o}chez} R.~A.,  {Harding} A.~K.,  2000, \mn@doi [\apj]
  {10.1086/317224}, \href
  {https://ui.adsabs.harvard.edu/abs/2000ApJ...544.1067A} {544, 1067}

\bibitem[\protect\citeauthoryear{{Arnaud}}{{Arnaud}}{1996}]{Arnaud1996}
{Arnaud} K.~A.,  1996, {XSPEC: The First Ten Years}.
p.~17

\bibitem[\protect\citeauthoryear{{Bailer-Jones}, {Rybizki}, {Fouesneau},
  {Mantelet}  \& {Andrae}}{{Bailer-Jones} et~al.}{2018}]{Balier-Jones18}
{Bailer-Jones} C.~A.~L.,  {Rybizki} J.,  {Fouesneau} M.,  {Mantelet} G.,
  {Andrae} R.,  2018, \mn@doi [\aj] {10.3847/1538-3881/aacb21}, \href
  {https://ui.adsabs.harvard.edu/abs/2018AJ....156...58B} {156, 58}

\bibitem[\protect\citeauthoryear{{Barlow}}{{Barlow}}{1989}]{Barlow89}
{Barlow} R.,  1989, {Statistics. A guide to the use of statistical methods in
  the physical sciences}

\bibitem[\protect\citeauthoryear{{Basko}}{{Basko}}{1978}]{Basko1978}
{Basko} M.~M.,  1978, \mn@doi [\apj] {10.1086/156260}, \href
  {https://ui.adsabs.harvard.edu/abs/1978ApJ...223..268B} {223, 268}

\bibitem[\protect\citeauthoryear{{Basko}}{{Basko}}{1980}]{Basko1980}
{Basko} M.~M.,  1980, \aap, \href
  {https://ui.adsabs.harvard.edu/abs/1980A&A....87..330B} {87, 330}

\bibitem[\protect\citeauthoryear{{Becker} et~al.,}{{Becker}
  et~al.}{2012}]{Becker12}
{Becker} P.~A.,  et~al., 2012, \mn@doi [\aap] {10.1051/0004-6361/201219065},
  \href {https://ui.adsabs.harvard.edu/abs/2012A&A...544A.123B} {544, A123}

\bibitem[\protect\citeauthoryear{{Beri} \& {Paul}}{{Beri} \&
  {Paul}}{2017}]{Beri17}
{Beri} A.,  {Paul} B.,  2017, \mn@doi [\na] {10.1016/j.newast.2017.05.001},
  \href {https://ui.adsabs.harvard.edu/abs/2017NewA...56...94B} {56, 94}

\bibitem[\protect\citeauthoryear{{Beri}, {Jain}, {Paul}  \& {Raichur}}{{Beri}
  et~al.}{2014}]{Beri14}
{Beri} A.,  {Jain} C.,  {Paul} B.,   {Raichur} H.,  2014, \mn@doi [\mnras]
  {10.1093/mnras/stu087}, \href
  {https://ui.adsabs.harvard.edu/abs/2014MNRAS.439.1940B} {439, 1940}

\bibitem[\protect\citeauthoryear{{Bildsten} et~al.,}{{Bildsten}
  et~al.}{1997}]{Bildsten1997}
{Bildsten} L.,  et~al., 1997, \mn@doi [\apjs] {10.1086/313060}, \href
  {https://ui.adsabs.harvard.edu/abs/1997ApJS..113..367B} {113, 367}

\bibitem[\protect\citeauthoryear{{Bissinger n{\'e} K{\"u}hnel}
  et~al.,}{{Bissinger n{\'e} K{\"u}hnel} et~al.}{2020}]{Bissingerne2020}
{Bissinger n{\'e} K{\"u}hnel} M.,  et~al., 2020, \mn@doi [\aap]
  {10.1051/0004-6361/201935666}, \href
  {https://ui.adsabs.harvard.edu/abs/2020A&A...634A..99B} {634, A99}

\bibitem[\protect\citeauthoryear{{Burderi}, {Di Salvo}, {Robba}, {La Barbera}
  \& {Guainazzi}}{{Burderi} et~al.}{2000}]{Burderi2000}
{Burderi} L.,  {Di Salvo} T.,  {Robba} N.~R.,  {La Barbera} A.,   {Guainazzi}
  M.,  2000, \mn@doi [\apj] {10.1086/308336}, \href
  {https://ui.adsabs.harvard.edu/abs/2000ApJ...530..429B} {530, 429}

\bibitem[\protect\citeauthoryear{{Chen}, {Qu}, {Zhang}, {Zhang}  \&
  {Zhang}}{{Chen} et~al.}{2008}]{Chen08}
{Chen} W.,  {Qu} J.-l.,  {Zhang} S.,  {Zhang} F.,   {Zhang} G.-b.,  2008,
  \mn@doi [\caa] {10.1016/j.chinastron.2008.07.012}, \href
  {https://ui.adsabs.harvard.edu/abs/2008ChA&A..32..241C} {32, 241}

\bibitem[\protect\citeauthoryear{{Coburn}, {Heindl}, {Rothschild}, {Gruber},
  {Kreykenbohm}, {Wilms}, {Kretschmar}  \& {Staubert}}{{Coburn}
  et~al.}{2002a}]{Coburn2002}
{Coburn} W.,  {Heindl} W.~A.,  {Rothschild} R.~E.,  {Gruber} D.~E.,
  {Kreykenbohm} I.,  {Wilms} J.,  {Kretschmar} P.,   {Staubert} R.,  2002a,
  \mn@doi [\apj] {10.1086/343033}, \href
  {https://ui.adsabs.harvard.edu/abs/2002ApJ...580..394C} {580, 394}

\bibitem[\protect\citeauthoryear{{Coburn}, {Heindl}, {Rothschild}, {Gruber},
  {Kreykenbohm}, {Wilms}, {Kretschmar}  \& {Staubert}}{{Coburn}
  et~al.}{2002b}]{Coburn02}
{Coburn} W.,  {Heindl} W.~A.,  {Rothschild} R.~E.,  {Gruber} D.~E.,
  {Kreykenbohm} I.,  {Wilms} J.,  {Kretschmar} P.,   {Staubert} R.,  2002b,
  \mn@doi [\apj] {10.1086/343033}, \href
  {https://ui.adsabs.harvard.edu/abs/2002ApJ...580..394C} {580, 394}

\bibitem[\protect\citeauthoryear{{Coley} et~al.,}{{Coley}
  et~al.}{2019}]{Coley19}
{Coley} J.~B.,  et~al., 2019, The Astronomer's Telegram, \href
  {https://ui.adsabs.harvard.edu/abs/2019ATel12684....1C} {12684, 1}

\bibitem[\protect\citeauthoryear{{Corbet}}{{Corbet}}{1986}]{Corbet86}
{Corbet} R.~H.~D.,  1986, \mn@doi [\mnras] {10.1093/mnras/220.4.1047}, \href
  {https://ui.adsabs.harvard.edu/abs/1986MNRAS.220.1047C} {220, 1047}

\bibitem[\protect\citeauthoryear{{Devasia}, {James}, {Paul}  \&
  {Indulekha}}{{Devasia} et~al.}{2011}]{Devasia11}
{Devasia} J.,  {James} M.,  {Paul} B.,   {Indulekha} K.,  2011, \mn@doi
  [\mnras] {10.1111/j.1365-2966.2011.18407.x}, \href
  {https://ui.adsabs.harvard.edu/abs/2011MNRAS.414.1023D} {414, 1023}

\bibitem[\protect\citeauthoryear{{Ebisawa}, {Day}, {Kallman}, {Nagase},
  {Kotani}, {Kawashima}, {Kitamoto}  \& {Woo}}{{Ebisawa}
  et~al.}{1996}]{Ebisawa1996}
{Ebisawa} K.,  {Day} C. S.~R.,  {Kallman} T.~R.,  {Nagase} F.,  {Kotani} T.,
  {Kawashima} K.,  {Kitamoto} S.,   {Woo} J.~W.,  1996, \mn@doi [\pasj]
  {10.1093/pasj/48.3.425}, \href
  {https://ui.adsabs.harvard.edu/abs/1996PASJ...48..425E} {48, 425}

\bibitem[\protect\citeauthoryear{Falkner}{Falkner}{2018}]{Falkner2018}
Falkner S.,  2018, doctoralthesis, Friedrich-Alexander-Universit{\"a}t
  Erlangen-N{\"u}rnberg (FAU)

\bibitem[\protect\citeauthoryear{{Ferrigno}, {Falanga}, {Bozzo}, {Becker},
  {Klochkov}  \& {Santangelo}}{{Ferrigno} et~al.}{2011}]{Ferrigno11}
{Ferrigno} C.,  {Falanga} M.,  {Bozzo} E.,  {Becker} P.~A.,  {Klochkov} D.,
  {Santangelo} A.,  2011, \mn@doi [\aap] {10.1051/0004-6361/201116826}, \href
  {https://ui.adsabs.harvard.edu/abs/2011A&A...532A..76F} {532, A76}

\bibitem[\protect\citeauthoryear{{Forman}, {Jones}  \& {Tananbaum}}{{Forman}
  et~al.}{1976}]{Forman76}
{Forman} W.,  {Jones} C.,   {Tananbaum} H.,  1976, \mn@doi [\apjl]
  {10.1086/182126}, \href
  {https://ui.adsabs.harvard.edu/abs/1976ApJ...206L..29F} {206, L29}

\bibitem[\protect\citeauthoryear{{Galloway}, {Wang}  \& {Morgan}}{{Galloway}
  et~al.}{2005}]{Galloway05}
{Galloway} D.~K.,  {Wang} Z.,   {Morgan} E.~H.,  2005, \mn@doi [\apj]
  {10.1086/497573}, \href
  {https://ui.adsabs.harvard.edu/abs/2005ApJ...635.1217G} {635, 1217}

\bibitem[\protect\citeauthoryear{{Harding}}{{Harding}}{1994}]{Harding94}
{Harding} A.~K.,  1994, in {Holt} S.,  {Day} C.~S.,  eds,  American Institute
  of Physics Conference Series Vol. 308, The Evolution of X-ray Binariese.
  p.~429, \mn@doi{10.1063/1.45983}

\bibitem[\protect\citeauthoryear{{Hemphill}, {Coley}, {Fuerst}, {Kretschmar},
  {Kuehnel}, {Malacaria}  \& {Pottschmidt}}{{Hemphill}
  et~al.}{2019}]{Hemphill19}
{Hemphill} P.,  {Coley} J.,  {Fuerst} F.,  {Kretschmar} P.,  {Kuehnel} M.,
  {Malacaria} C.,   {Pottschmidt} K.,  2019, The Astronomer's Telegram, \href
  {https://ui.adsabs.harvard.edu/abs/2019ATel12556....1H} {12556, 1}

\bibitem[\protect\citeauthoryear{{Jaisawal} et~al.,}{{Jaisawal}
  et~al.}{2019}]{Jaisawal19}
{Jaisawal} G.~K.,  et~al., 2019, The Astronomer's Telegram, \href
  {https://ui.adsabs.harvard.edu/abs/2019ATel12515....1J} {12515, 1}

\bibitem[\protect\citeauthoryear{{James}, {Paul}, {Devasia}  \&
  {Indulekha}}{{James} et~al.}{2011}]{James11}
{James} M.,  {Paul} B.,  {Devasia} J.,   {Indulekha} K.,  2011, \mn@doi
  [\mnras] {10.1111/j.1365-2966.2010.17543.x}, \href
  {https://ui.adsabs.harvard.edu/abs/2011MNRAS.410.1489J} {410, 1489}

\bibitem[\protect\citeauthoryear{{Jenke} \& {Finger}}{{Jenke} \&
  {Finger}}{2011}]{Jenke11}
{Jenke} P.,  {Finger} M.~H.,  2011, The Astronomer's Telegram, \href
  {https://ui.adsabs.harvard.edu/abs/2011ATel.3839....1J} {3839, 1}

\bibitem[\protect\citeauthoryear{{Ji} et~al.,}{{Ji} et~al.}{2020}]{Ji2020}
{Ji} L.,  et~al., 2020, \mn@doi [\mnras] {10.1093/mnras/staa569}, \href
  {https://ui.adsabs.harvard.edu/abs/2020MNRAS.493.5680J} {493, 5680}

\bibitem[\protect\citeauthoryear{{Kraus}, {Nollert}, {Ruder}  \&
  {Riffert}}{{Kraus} et~al.}{1995}]{Kraus95}
{Kraus} U.,  {Nollert} H.~P.,  {Ruder} H.,   {Riffert} H.,  1995, \mn@doi
  [\apj] {10.1086/176182}, \href
  {https://ui.adsabs.harvard.edu/abs/1995ApJ...450..763K} {450, 763}

\bibitem[\protect\citeauthoryear{{Kraus}, {Blum}, {Schulte}, {Ruder}  \&
  {Meszaros}}{{Kraus} et~al.}{1996}]{Kraus96}
{Kraus} U.,  {Blum} S.,  {Schulte} J.,  {Ruder} H.,   {Meszaros} P.,  1996,
  \mn@doi [\apj] {10.1086/177653}, \href
  {https://ui.adsabs.harvard.edu/abs/1996ApJ...467..794K} {467, 794}

\bibitem[\protect\citeauthoryear{{Kreykenbohm}, {Kretschmar}, {Wilms},
  {Staubert}, {Kendziorra}, {Gruber}, {Heindl}  \& {Rothschild}}{{Kreykenbohm}
  et~al.}{1999}]{Kreykenbohm99}
{Kreykenbohm} I.,  {Kretschmar} P.,  {Wilms} J.,  {Staubert} R.,  {Kendziorra}
  E.,  {Gruber} D.~E.,  {Heindl} W.~A.,   {Rothschild} R.~E.,  1999, \aap,
  \href {https://ui.adsabs.harvard.edu/abs/1999A&A...341..141K} {341, 141}

\bibitem[\protect\citeauthoryear{{Lei}, {Chen}, {Qu}, {Song}, {Zhang}, {Lu},
  {Zhang}  \& {Li}}{{Lei} et~al.}{2009}]{Lei09}
{Lei} Y.-J.,  {Chen} W.,  {Qu} J.-L.,  {Song} L.-M.,  {Zhang} S.,  {Lu} Y.,
  {Zhang} H.-T.,   {Li} T.-P.,  2009, \mn@doi [\apj]
  {10.1088/0004-637X/707/2/1016}, \href
  {https://ui.adsabs.harvard.edu/abs/2009ApJ...707.1016L} {707, 1016}

\bibitem[\protect\citeauthoryear{{Lutovinov}, {Tsygankov}, {Suleimanov},
  {Mushtukov}, {Doroshenko}, {Nagirner}  \& {Poutanen}}{{Lutovinov}
  et~al.}{2015}]{Lutovinov15}
{Lutovinov} A.~A.,  {Tsygankov} S.~S.,  {Suleimanov} V.~F.,  {Mushtukov} A.~A.,
   {Doroshenko} V.,  {Nagirner} D.~I.,   {Poutanen} J.,  2015, \mn@doi [\mnras]
  {10.1093/mnras/stv125}, \href
  {https://ui.adsabs.harvard.edu/abs/2015MNRAS.448.2175L} {448, 2175}

\bibitem[\protect\citeauthoryear{Maitra}{Maitra}{2017}]{Maitra17b}
Maitra C.,  2017, \mn@doi [J. Astrophys. Astron.] {10.1007/s12036-017-9476-3},
  38, 50

\bibitem[\protect\citeauthoryear{Maitra \& Paul}{Maitra \&
  Paul}{2013}]{Maitra_2013}
Maitra C.,  Paul B.,  2013, \mn@doi [The Astrophysical Journal]
  {10.1088/0004-637x/763/2/79}, 763, 79

\bibitem[\protect\citeauthoryear{{Maitra}, {Paul}  \& {Naik}}{{Maitra}
  et~al.}{2012}]{Maitra12}
{Maitra} C.,  {Paul} B.,   {Naik} S.,  2012, \mn@doi [\mnras]
  {10.1111/j.1365-2966.2011.20196.x}, \href
  {https://ui.adsabs.harvard.edu/abs/2012MNRAS.420.2307M} {420, 2307}

\bibitem[\protect\citeauthoryear{{Maitra}, {Raichur}, {Pradhan}  \&
  {Paul}}{{Maitra} et~al.}{2017}]{Maitra17a}
{Maitra} C.,  {Raichur} H.,  {Pradhan} P.,   {Paul} B.,  2017, \mn@doi [\mnras]
  {10.1093/mnras/stx1281}, \href
  {https://ui.adsabs.harvard.edu/abs/2017MNRAS.470..713M} {470, 713}

\bibitem[\protect\citeauthoryear{{McCollum} \& {Laine}}{{McCollum} \&
  {Laine}}{2019}]{McCollum19}
{McCollum} B.,  {Laine} S.,  2019, The Astronomer's Telegram, \href
  {https://ui.adsabs.harvard.edu/abs/2019ATel12560....1M} {12560, 1}

\bibitem[\protect\citeauthoryear{{Mereminskiy}, {Lutovinov}, {Tsygankov},
  {Semena}  \& {Shtykovskiy}}{{Mereminskiy} et~al.}{2019}]{Mereminskiy19}
{Mereminskiy} I.~A.,  {Lutovinov} A.~A.,  {Tsygankov} S.~S.,  {Semena} A.~N.,
  {Shtykovskiy} A.~E.,  2019, The Astronomer's Telegram, \href
  {https://ui.adsabs.harvard.edu/abs/2019ATel12514....1M} {12514, 1}

\bibitem[\protect\citeauthoryear{{Meszaros} \& {Nagel}}{{Meszaros} \&
  {Nagel}}{1985}]{Meszaros85}
{Meszaros} P.,  {Nagel} W.,  1985, \mn@doi [\apj] {10.1086/163594}, \href
  {https://ui.adsabs.harvard.edu/abs/1985ApJ...298..147M} {298, 147}

\bibitem[\protect\citeauthoryear{{Molkov}, {Lutovinov}  \& {Grebenev}}{{Molkov}
  et~al.}{2003}]{Molkov03}
{Molkov} S.,  {Lutovinov} A.,   {Grebenev} S.,  2003, \mn@doi [\aap]
  {10.1051/0004-6361:20031481}, \href
  {https://ui.adsabs.harvard.edu/abs/2003A&A...411L.357M} {411, L357}

\bibitem[\protect\citeauthoryear{{Molkov}, {Lutovinov}, {Tsygankov},
  {Mereminskiy}  \& {Mushtukov}}{{Molkov} et~al.}{2019}]{Molkov19}
{Molkov} S.,  {Lutovinov} A.,  {Tsygankov} S.,  {Mereminskiy} I.,   {Mushtukov}
  A.,  2019, \mn@doi [\apjl] {10.3847/2041-8213/ab3e4d}, \href
  {https://ui.adsabs.harvard.edu/abs/2019ApJ...883L..11M} {883, L11}

\bibitem[\protect\citeauthoryear{{Mushtukov}, {Suleimanov}, {Tsygankov}  \&
  {Poutanen}}{{Mushtukov} et~al.}{2015a}]{Mushtukov2015}
{Mushtukov} A.~A.,  {Suleimanov} V.~F.,  {Tsygankov} S.~S.,   {Poutanen} J.,
  2015a, \mn@doi [\mnras] {10.1093/mnras/stu2484}, \href
  {https://ui.adsabs.harvard.edu/abs/2015MNRAS.447.1847M} {447, 1847}

\bibitem[\protect\citeauthoryear{{Mushtukov}, {Suleimanov}, {Tsygankov}  \&
  {Poutanen}}{{Mushtukov} et~al.}{2015b}]{Mushtukov15}
{Mushtukov} A.~A.,  {Suleimanov} V.~F.,  {Tsygankov} S.~S.,   {Poutanen} J.,
  2015b, \mn@doi [\mnras] {10.1093/mnras/stu2484}, \href
  {https://ui.adsabs.harvard.edu/abs/2015MNRAS.447.1847M} {447, 1847}

\bibitem[\protect\citeauthoryear{{Nagase}}{{Nagase}}{1989}]{Nagase89}
{Nagase} F.,  1989, \pasj, \href
  {https://ui.adsabs.harvard.edu/abs/1989PASJ...41....1N} {41, 1}

\bibitem[\protect\citeauthoryear{{Nagel}}{{Nagel}}{1981}]{Nagel81}
{Nagel} W.,  1981, \mn@doi [\apj] {10.1086/159464}, \href
  {https://ui.adsabs.harvard.edu/abs/1981ApJ...251..288N} {251, 288}

\bibitem[\protect\citeauthoryear{{Nakajima} et~al.,}{{Nakajima}
  et~al.}{2019}]{Nakajima19}
{Nakajima} M.,  et~al., 2019, The Astronomer's Telegram, \href
  {https://ui.adsabs.harvard.edu/abs/2019ATel12498....1N} {12498, 1}

\bibitem[\protect\citeauthoryear{{Nishimura}}{{Nishimura}}{2014}]{Nishimura14}
{Nishimura} O.,  2014, \mn@doi [\apj] {10.1088/0004-637X/781/1/30}, \href
  {https://ui.adsabs.harvard.edu/abs/2014ApJ...781...30N} {781, 30}

\bibitem[\protect\citeauthoryear{{Orlandini}, {Frontera}, {Masetti}, {Sguera}
  \& {Sidoli}}{{Orlandini} et~al.}{2012}]{Orlandini12}
{Orlandini} M.,  {Frontera} F.,  {Masetti} N.,  {Sguera} V.,   {Sidoli} L.,
  2012, \mn@doi [\apj] {10.1088/0004-637X/748/2/86}, \href
  {https://ui.adsabs.harvard.edu/abs/2012ApJ...748...86O} {748, 86}

\bibitem[\protect\citeauthoryear{{Parmar}, {White}  \& {Stella}}{{Parmar}
  et~al.}{1989}]{Parmar89}
{Parmar} A.~N.,  {White} N.~E.,   {Stella} L.,  1989, \mn@doi [\apj]
  {10.1086/167205}, \href
  {https://ui.adsabs.harvard.edu/abs/1989ApJ...338..373P} {338, 373}

\bibitem[\protect\citeauthoryear{{Poutanen}, {Mushtukov}, {Suleimanov},
  {Tsygankov}, {Nagirner}, {Doroshenko}  \& {Lutovinov}}{{Poutanen}
  et~al.}{2013}]{Poutanen13}
{Poutanen} J.,  {Mushtukov} A.~A.,  {Suleimanov} V.~F.,  {Tsygankov} S.~S.,
  {Nagirner} D.~I.,  {Doroshenko} V.,   {Lutovinov} A. e.~A.,  2013, \mn@doi
  [\apj] {10.1088/0004-637X/777/2/115}, \href
  {https://ui.adsabs.harvard.edu/abs/2013ApJ...777..115P} {777, 115}

\bibitem[\protect\citeauthoryear{{Priedhorsky} \& {Terrell}}{{Priedhorsky} \&
  {Terrell}}{1984}]{Priedhorsky84}
{Priedhorsky} W.~C.,  {Terrell} J.,  1984, \mn@doi [\apj] {10.1086/162039},
  \href {https://ui.adsabs.harvard.edu/abs/1984ApJ...280..661P} {280, 661}

\bibitem[\protect\citeauthoryear{{Reig}}{{Reig}}{2011}]{Reig11}
{Reig} P.,  2011, \mn@doi [\apss] {10.1007/s10509-010-0575-8}, \href
  {https://ui.adsabs.harvard.edu/abs/2011Ap&SS.332....1R} {332, 1}

\bibitem[\protect\citeauthoryear{{Reig} \& {Milonaki}}{{Reig} \&
  {Milonaki}}{2016}]{Reig16}
{Reig} P.,  {Milonaki} F.,  2016, \mn@doi [\aap] {10.1051/0004-6361/201629200},
  \href {https://ui.adsabs.harvard.edu/abs/2016A&A...594A..45R} {594, A45}

\bibitem[\protect\citeauthoryear{{Reig} \& {Nespoli}}{{Reig} \&
  {Nespoli}}{2013}]{Reig13}
{Reig} P.,  {Nespoli} E.,  2013, \mn@doi [\aap] {10.1051/0004-6361/201219806},
  \href {https://ui.adsabs.harvard.edu/abs/2013A&A...551A...1R} {551, A1}

\bibitem[\protect\citeauthoryear{{Sch{\"o}nherr}, {Wilms}, {Kretschmar},
  {Kreykenbohm}, {Santangelo}, {Rothschild}, {Coburn}  \&
  {Staubert}}{{Sch{\"o}nherr} et~al.}{2007}]{Schonherr07}
{Sch{\"o}nherr} G.,  {Wilms} J.,  {Kretschmar} P.,  {Kreykenbohm} I.,
  {Santangelo} A.,  {Rothschild} R.~E.,  {Coburn} W.,   {Staubert} R.,  2007,
  \mn@doi [\aap] {10.1051/0004-6361:20077218}, \href
  {https://ui.adsabs.harvard.edu/abs/2007A&A...472..353S} {472, 353}

\bibitem[\protect\citeauthoryear{{Sch{\"o}nherr} et~al.,}{{Sch{\"o}nherr}
  et~al.}{2014}]{Schonherr14}
{Sch{\"o}nherr} G.,  et~al., 2014, \mn@doi [\aap]
  {10.1051/0004-6361/201322448}, \href
  {https://ui.adsabs.harvard.edu/abs/2014A&A...564L...8S} {564, L8}

\bibitem[\protect\citeauthoryear{{Sootome} et~al.,}{{Sootome}
  et~al.}{2011}]{Sootome11}
{Sootome} T.,  et~al., 2011, The Astronomer's Telegram, \href
  {https://ui.adsabs.harvard.edu/abs/2011ATel.3829....1S} {3829, 1}

\bibitem[\protect\citeauthoryear{{Staubert} et~al.,}{{Staubert}
  et~al.}{2019}]{Staubert2019}
{Staubert} R.,  et~al., 2019, \mn@doi [\aap] {10.1051/0004-6361/201834479},
  \href {https://ui.adsabs.harvard.edu/abs/2019A&A...622A..61S} {622, A61}

\bibitem[\protect\citeauthoryear{{Strader}, {Chomiuk}, {Swihart}  \&
  {Aydi}}{{Strader} et~al.}{2019}]{Strader19}
{Strader} J.,  {Chomiuk} L.,  {Swihart} S.,   {Aydi} E.,  2019, The
  Astronomer's Telegram, \href
  {https://ui.adsabs.harvard.edu/abs/2019ATel12554....1S} {12554, 1}

\bibitem[\protect\citeauthoryear{{Tanaka}}{{Tanaka}}{1986}]{Tanaka1986}
{Tanaka} Y.,  1986, {Observations of Compact X-Ray Sources}.
p.~198, \mn@doi{10.1007/3-540-16764-1_12}

\bibitem[\protect\citeauthoryear{{Titarchuk}}{{Titarchuk}}{1994}]{Titarchuk94}
{Titarchuk} L.,  1994, \mn@doi [\apj] {10.1086/174760}, \href
  {https://ui.adsabs.harvard.edu/abs/1994ApJ...434..570T} {434, 570}

\bibitem[\protect\citeauthoryear{{Tsygankov}, {Lutovinov}, {Churazov}  \&
  {Sunyaev}}{{Tsygankov} et~al.}{2006}]{Tsygankov06}
{Tsygankov} S.~S.,  {Lutovinov} A.~A.,  {Churazov} E.~M.,   {Sunyaev} R.~A.,
  2006, \mn@doi [\mnras] {10.1111/j.1365-2966.2006.10610.x}, \href
  {https://ui.adsabs.harvard.edu/abs/2006MNRAS.371...19T} {371, 19}

\bibitem[\protect\citeauthoryear{{Tsygankov}, {Lutovinov}, {Churazov}  \&
  {Sunyaev}}{{Tsygankov} et~al.}{2007}]{Tsygankov07}
{Tsygankov} S.~S.,  {Lutovinov} A.~A.,  {Churazov} E.~M.,   {Sunyaev} R.~A.,
  2007, \mn@doi [Astronomy Letters] {10.1134/S1063773707060023}, \href
  {https://ui.adsabs.harvard.edu/abs/2007AstL...33..368T} {33, 368}

\bibitem[\protect\citeauthoryear{{Tuo} et~al.,}{{Tuo} et~al.}{2020}]{Tuo2020}
{Tuo} Y.~L.,  et~al., 2020, arXiv e-prints, \href
  {https://ui.adsabs.harvard.edu/abs/2020arXiv200413307T} {p. arXiv:2004.13307}

\bibitem[\protect\citeauthoryear{{Verner}, {Ferland}, {Korista}  \&
  {Yakovlev}}{{Verner} et~al.}{1996}]{Verner1996}
{Verner} D.~A.,  {Ferland} G.~J.,  {Korista} K.~T.,   {Yakovlev} D.~G.,  1996,
  \mn@doi [\apj] {10.1086/177435}, \href
  {https://ui.adsabs.harvard.edu/abs/1996ApJ...465..487V} {465, 487}

\bibitem[\protect\citeauthoryear{{Wilms}, {Allen}  \& {McCray}}{{Wilms}
  et~al.}{2000}]{Wilms2000}
{Wilms} J.,  {Allen} A.,   {McCray} R.,  2000, \mn@doi [\apj] {10.1086/317016},
  \href {https://ui.adsabs.harvard.edu/abs/2000ApJ...542..914W} {542, 914}

\makeatother
\end{thebibliography}

\label{lastpage}
\end{document}